%% file: MLG-23-003_temp.tex
\documentclass[11pt,twoside,a4paper,cmspaper,final,collab]{cms-tdr}
\def\svnVersion{3edb533}\def\svnDate{2026/08/26}\def\cmsCernNoTag{CERN-EP-2025-092}\def\cmsCernDate{\today}\def\cmsMessage{Published in Machine Learning: Science and Technology as \href{http://dx.doi.org/10.1088/2632-2153/ae8c11}{\doi{10.1088/2632-2153/ae8c11}.}}
\begin{document}\cmsNoteHeader{MLG-23-003}

\ifthenelse{\boolean{cms@external}}{\providecommand{\cmsLeft}{upper\xspace}}{\providecommand{\cmsLeft}{left\xspace}}
\ifthenelse{\boolean{cms@external}}{\providecommand{\cmsRight}{lower\xspace}}{\providecommand{\cmsRight}{right\xspace}}
\ifthenelse{\boolean{cms@external}}{\providecommand{\cmsUpperLeft}{upper\xspace}}{\providecommand{\cmsUpperLeft}{upper left\xspace}}
\ifthenelse{\boolean{cms@external}}{\providecommand{\cmsUpperRight}{middle\xspace}}{\providecommand{\cmsUpperRight}{upper right\xspace}}

\newcommand{\njets}{\ensuremath{N_\text{jets}}\xspace}
\newcommand{\sone}{\ensuremath{S_{1}}\xspace}
\newcommand{\stwo}{\ensuremath{S_{2}}\xspace}
\newcommand{\sonetwo}{\ensuremath{S_{1,2}}\xspace}
\newcommand{\sonei}{\ensuremath{S_{1,i}}\xspace}
\newcommand{\stwoi}{\ensuremath{S_{2,i}}\xspace}
\newcommand{\aone}{\ensuremath{a_{1}}\xspace}
\newcommand{\atwo}{\ensuremath{a_{2}}\xspace}
\newcommand{\aonetwo}{\ensuremath{a_{1,2}}\xspace}
\newcommand{\bone}{\ensuremath{b_{1}}\xspace}
\newcommand{\btwo}{\ensuremath{b_{2}}\xspace}
\newcommand{\bonetwo}{\ensuremath{b_{1,2}}\xspace}
\newcommand{\snni}[1]  {\ensuremath{S^\text{NN}_{#1}}\xspace}
\newcommand{\mstop}{\ensuremath{m_{\PSQt}}\xspace}
\newcommand{\ttjets}{\ensuremath{\ttbar + \text{jets}}\xspace}

\newcommand{\Areg}{\ensuremath{\text{A}}\xspace}
\newcommand{\Breg}{\ensuremath{\text{B}}\xspace}
\newcommand{\Creg}{\ensuremath{\text{C}}\xspace}
\newcommand{\Dreg}{\ensuremath{\text{D}}\xspace}

\newcommand{\Ni}{\ensuremath{N_{i}}\xspace}
\newcommand{\NAreg}{\ensuremath{N_{\Areg}}\xspace}
\newcommand{\NApred}{\ensuremath{N_{\Areg}^{\text{pred}}}\xspace}
\newcommand{\NBreg}{\ensuremath{N_{\Breg}}\xspace}
\newcommand{\NCreg}{\ensuremath{N_{\Creg}}\xspace}
\newcommand{\NDreg}{\ensuremath{N_{\Dreg}}\xspace}

\newcommand{\ytrue}{\ensuremath{y_{\text{true}}}\xspace}
\newcommand{\ypredi}{\ensuremath{y^{(i)}_{\text{pred}}}\xspace}
\newcommand{\LTotal}{\ensuremath{L_{\text{total}}}\xspace}
\newcommand{\LMR}{\ensuremath{L_{\text{MR}}}\xspace}
\newcommand{\LBCE}{\ensuremath{L_{\text{BCE}}}\xspace}
\newcommand{\LDisCo}{\ensuremath{L_{\text{DisCo}}}\xspace}
\newcommand{\LClosure}{\ensuremath{L_{\text{nonclosure}}}\xspace}
\newcommand{\lamClosure}{\ensuremath{\lambda_{\text{nonclosure}}}\xspace}
\newcommand{\lamDisCo}{\ensuremath{\lambda_{\text{DisCo}}}\xspace}
\newcommand{\epsClosure}{\ensuremath{\epsilon_{\text{nonclosure}}}\xspace}
\newcommand{\epsDisCo}{\ensuremath{\epsilon_{\text{DisCo}}}\xspace}

\newcommand{\disco}{ABCDisCo\xspace}
\newcommand{\discotec}{ABCDisCoTEC\xspace}
\newcommand{\nonclosure}{\ensuremath{{C}\hspace{-0.70em}/\kern 0.0em}\xspace}

\newcommand{\mpred}{\ensuremath{m_{\text{pred}}}\xspace}
\newcommand{\mtrue}{\ensuremath{m_{\text{true}}}\xspace}

\newcommand{\Nsig}{\ensuremath{N_{\text{sig}}}\xspace}
\newcommand{\Nbkg}{\ensuremath{N_{\text{bkg}}}\xspace}
\newcommand{\Ntot}{\ensuremath{N_{\text{total}}}\xspace}

\newcommand{\ADAM}{\textsc{Adam}\xspace}
\newcommand{\PYTORCH}{\textsc{PyTorch}\xspace}

\newcommand{\Fig}[1]{Fig.~\ref{fig:#1}}
\newcommand{\Equ}[1]{Eq.~\eqref{eq:#1}}
\newcommand{\Sec}[1]{Section~\ref{sec:#1}}
\newcommand{\Secs}[2]{Sections~\ref{sec:#1} and \ref{sec:#2}}

\newcommand{\Figur}[1]{Figure~\ref{fig:#1}}
\newcommand{\Equation}[1]{Equation~\eqref{eq:#1}}

\cmsNoteHeader{MLG-23-003}
\title{Machine learning method for enforcing variable independence in background estimation with LHC data: ABCDisCoTEC}
\titlerunning{Machine learning method for enforcing variable independence in background estimation with LHC data}
\date{\today}

\abstract{
A novel solution is presented for the problem of estimating the backgrounds of a signal search using observed data while simultaneously maximizing the sensitivity of the search to the signal. The ``ABCD method'' provides a reliable framework for background estimation by partitioning events into one signal-enhanced region (A) and three background-enhanced control regions (B, C, and D) via two smoothly varying, statistically independent variables. In practice, even slight correlations between the two variables can significantly undermine the method's performance. Thus, choosing appropriate variables by hand can present a formidable challenge, especially when background and signal differ only subtly. To address this issue, the ABCD with distance correlation (ABCDisCo) method was developed to construct two learned variables via a neural network trained to provide strong signal-background discrimination with small values of the distance correlation measure between the two learned variables. However, relying solely on minimizing the distance correlation can result in learned variables that may not have distributions of background events that are smoothly varying and localized at extreme values, as necessary for the validity of the background estimation. The ABCDisCo training enhanced with closure (ABCDisCoTEC) method is introduced to solve this issue by directly minimizing the nonclosure, expressed as a dedicated differentiable loss term. This extended method is applied to a data set of proton-proton collisions at a center-of-mass energy of 13\TeV recorded by the CMS detector at the CERN LHC. Additionally, given the complexity of the minimization problem with constraints on multiple loss terms, the modified differential method of multipliers is applied and shown to greatly improve the stability and robustness of the ABCDisCoTEC method, compared to grid search hyperparameter optimization procedures.
}

\hypersetup{%
pdfauthor={CMS Collaboration},%
pdftitle={Machine learning method for enforcing variable independence in background estimation with LHC data},
pdfsubject={CMS},%
pdfkeywords={CMS, supersymmetry, SUSY, neural network}}

\maketitle 

\section{Introduction}
\label{sec:intro}

Particle physics experiments generally analyze complex data sets to make precision measurements within the framework of the standard model (SM) or to search for new phenomena beyond the SM.
At high-luminosity proton-proton colliders like the Large Hadron Collider (LHC), these data sets consist of the information associated with the outgoing particles produced in each independent collision of proton bunches as recorded with a particle detector; the data associated with each collision is referred to as an ``event.'' 
These analyses require separation of the events of interest (``signal events'') from all other events (``background events''). 
A wide range of techniques has been developed for robust estimation of the background for each analysis.
Relying solely on simulated data to estimate backgrounds can result in large systematic uncertainties associated with the nature of the modeling. This limits the overall power of the measurement.
Therefore, it is often preferable to use observed data to estimate the background, avoiding potential modeling discrepancies.
Additionally, for data sets in which differences between signal and background in traditional physical variables are subtle, such variables are generally not sufficient for effective discrimination.
Signal sensitivity improvements have been achieved by training neural networks (NNs) that can more effectively separate signal from background by taking advantage of complex correlations among different variables.

A commonly used technique known as the ``ABCD method''~\cite{D0:1994wmk,Behnke:2013pga,Buttinger:2018} provides a reliable background estimation by using two statistically independent variables to define control regions (CRs) for extrapolation into a signal region (SR).
However, finding two independent variables that can be successfully leveraged for the ABCD method is often challenging.
If the chosen variables are not independent, the performance of the method worsens, as quantified by the difference between the prediction and the true value, referred to as nonclosure.
To address this problem, Ref.~\cite{Kasieczka:2020pil} introduced the \disco method, which automates the creation of two uncorrelated NN outputs that can then be used for a robust background estimate.
The method employs distance correlation (DisCo)~\cite{Gabor:2007}, a statistical metric that captures a wide variety of both linear and nonlinear correlations,
to ensure that the two outputs are independent and concentrate the signal events in the SR, with limited contamination in adjacent CRs.

The two uncorrelated \disco outputs meet the conditions necessary for robust background estimation using the ABCD method.
However, the performance of the ABCD method cannot be directly targeted when training the \disco network with gradient descent optimization methods, as the nonclosure conditions are inherently nondifferentiable.
This paper presents a novel solution called \discotec (\disco training enhanced with closure), which directly minimizes the nonclosure to ensure that the background estimate is accurate.
We apply this new method to a search for new physics in high-energy proton-proton collision data recorded by the CMS detector~\cite{CMS:2008xjf} at the CERN LHC~\cite{Evans:2008zzb}, the first application of such a method to observed collider data.
Here, we focus on the generalizable aspects of the application; the specific details and results from the search are documented in Ref.~\cite{SUS-23-001}.  
As the \discotec model has a complex multi-term loss function, this paper also demonstrates the use of the modified differential method of multipliers (MDMM)~\cite{Platt:1987} to obtain an optimal NN training, which is the first application of MDMM with collider data.
While a reasonable training can be achieved using a manual optimization of the choice of hyperparameters---as is the case for the models used in Ref.~\cite{SUS-23-001}---MDMM reduces the number of training iterations necessary to converge on a final solution. 
The following sections provide additional details about the methods and the validation of the results.

\section{Methods}
\label{sec:methods}

\subsection{The ABCD method}
\label{sec:ABCD}

\begin{figure*}[htp]
   \centering
   \includegraphics[width=0.6\textwidth]{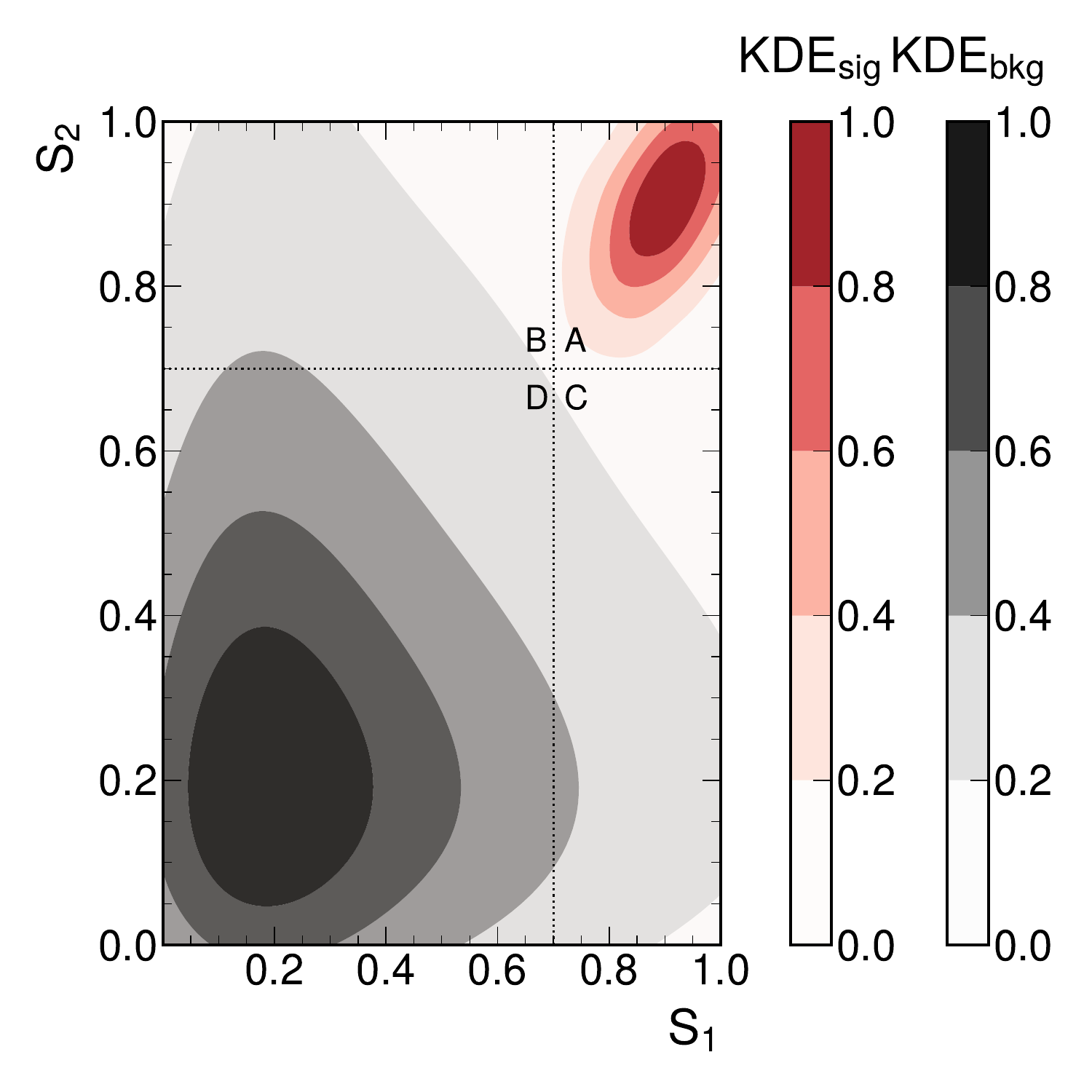}
   \caption{Schematic illustration of idealized signal (red) and background (grey) distributions in the ABCD plane, represented as Gaussian kernel density estimators (KDEs).
   \label{fig:ABCDCartoon}}
\end{figure*}

The ABCD method proceeds by defining four regions based on the intersection of two boundaries in statistically independent variables (referred to henceforth as discriminants, or \sone and \stwo), such that the signal and background are effectively separated.
Without loss of generality, we define the SR, labeled \Areg, to contain the events above the boundary in each variable, corresponding to the upper right-hand corner of the plane spanned by \sone and \stwo.
The adjacent \Breg, \Creg, and \Dreg regions serve as CRs, with relatively more background and fewer signal events.
In the ideal case, background is tightly concentrated in the \Dreg region in the lower left-hand corner, and signal resides almost exclusively in the \Areg region (upper right-hand corner), as illustrated in \Fig{ABCDCartoon}.

The prediction of the number of background events in the SR \Areg, \NApred, is calculated from the numbers of events in the surrounding CRs:
\begin{equation}
\label{eq:abcdRelation}
    \NApred = \frac{\NBreg\NCreg}{\NDreg},
\end{equation}
where \NBreg, \NCreg, and \NDreg correspond to the number of events in the \Breg, \Creg, and \Dreg CRs, respectively. 
In some situations, the ABCD space may be focused on one or a subset of background processes.
In this case, the \Ni counts are modified by subtracting the contributions of other backgrounds as estimated by simulation. 
If the two variables defining the ABCD plane are independent for background events, then $\NApred \approx \NAreg$, where \NAreg is the actual number of background events in the SR.
The variables are not required to be independent for signal events, as indicated by the positive covariance of the signal distribution in \Fig{ABCDCartoon}.

The event yield \Ni for each of the four regions is treated as a Poisson distribution in the underlying statistical model.
Thus, \Equ{abcdRelation} represents a constraint on the likelihood function of the statistical model, which is the product of the Poisson likelihoods for each region~\cite{Behnke:2013pga, Buttinger:2018}.
In this way, a maximum likelihood estimate for \NApred is computed when fitting the statistical model to the observed data.

A useful metric to quantify the accuracy of the prediction is the relative difference between the observed and predicted numbers of events in region \Areg in simulation.
This quantity, known as the nonclosure (\nonclosure), is computed as:
\begin{equation}
    \label{eq:nonclosure}
    \nonclosure = \left| 1 - \frac{\NBreg\NCreg}{\NAreg\NDreg} \right|.
\end{equation}

\Equation{nonclosure} shows that the nonclosure metric is computed from event counts above and below the boundaries; therefore, its values may be discontinuous.
This presents a significant challenge to train a machine learning model to minimize the nonclosure, as such models rely on the differentiability of the loss function.
In \Sec{model}, this issue is discussed in more detail, and a novel solution is presented.

\begin{figure*}[htp]
   \centering
   \includegraphics[width=0.8\textwidth]{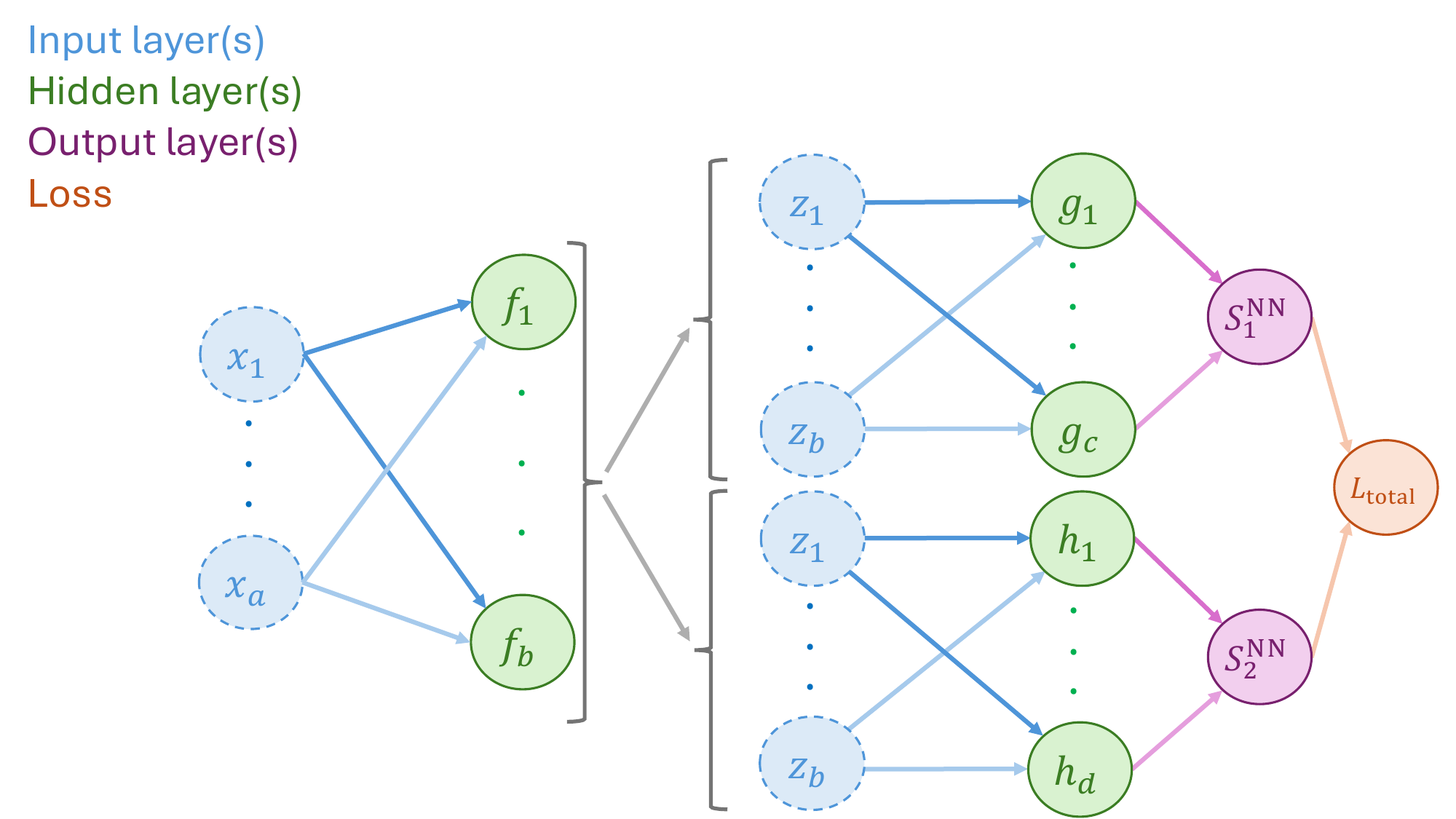}
   \caption{Diagrammatic layout of the \discotec NN model. 
    The $a$ input features $x_i$ are initially fed into a hidden layer $f$ with $b$ nodes. 
    The resulting output of this hidden layer, $z$, is fed into two binary classifiers, each with one hidden layer $g$ or $h$, with $c$ or $d$ nodes, respectively, and output layer \snni{1} or \snni{2}.
    Both outputs are used to compute the loss function \LTotal.
    Colored arrows indicate the direction of propagation from inputs to final network output.
   \label{fig:modelSchematic}}
\end{figure*}

The ABCD relationship is valid under two conditions: the independence of the two variables and low signal contamination in the \Breg, \Creg, and \Dreg CRs.
While the latter condition is not a strict requirement because the method can account for it, signal contamination still results in reduced signal identification efficiency
and therefore decreases the sensitivity of the analysis.
The primary goal of optimizing the performance of the ABCD method amounts to ensuring these two conditions are satisfied.
A more detailed description of the ABCD method is presented in Ref.~\cite{Kasieczka:2020pil}.

A key component of any background estimation method is to define validation regions (VRs) with minimal signal acceptance in which the validity and performance of the method can be evaluated.
The success of the VRs often depends on how closely they mirror fundamental kinematic properties of the SR.
In an ideal scenario, there is a fully independent VR to assess the performance and nonclosure of the ABCD method. 
However, in the absence of such a VR, we present an alternate approach in \Sec{data_validation}.

\subsection{The \discotec method}
\label{sec:model}

To achieve good performance from the ABCD method with two NN outputs, there are two primary objectives to satisfy.
First, the model must generate two binary classifier outputs (\snni{1}, \snni{2}) that adequately distinguish signal events from background events.
Second, the two classifier outputs need to be independent.
These two main criteria motivate the overall structure of the \discotec NN model and the form of the loss functions used during training.
As stated in Theorem 3 of Ref.~\cite{Gabor:2007}, a DisCo value of zero between two sets of variables implies independence.
Thus, minimization of DisCo can enforce independence between two observables and is therefore a useful metric for quantifying the feasibility of an ABCD background estimation.
As indicated in \Equ{totalLoss} below, we include both DisCo and nonclosure terms in the loss function. 

\Figur{modelSchematic} shows a diagrammatic representation of the NN model used to accomplish the two aforementioned tasks.
The core components are the dual, identically configured, binary classifiers that each generate a unique discriminant value between 0 and 1 for each event, corresponding to the NN's confidence that a given event should be classified as signal.

The minimal loss function for the ABCDisCoTEC model is composed of three distinct terms, which are explained in the following paragraphs:
\begin{equation}
\label{eq:totalLoss}
\begin{split}
    \LTotal =~& \lambda_{\text{BCE}} \left(\LBCE^{(1)} + \LBCE^{(2)}\right) \\
	& + \lamDisCo \LDisCo + \lamClosure \LClosure,
\end{split}
\end{equation}
where the hyperparameters $\lambda$ are positive multiplicative constants that control the relative importance of each loss term.
All terms in the loss function are jointly optimized during training.
This loss function can be extended (as discussed in \Sec{traindetails}) to include additional components that result in specific training outcomes.  
An example of such an extension could be a term that mitigates signal contamination in the CRs. 
However, additional loss terms increase the training complexity and without proper optimization may result in poor NN performance.

The first term in \Equ{totalLoss} represents two binary cross-entropy (BCE) terms, one for each binary classifier $i$; each takes the form:
\begin{equation}
\label{eq:binaryCrossEntropy}
\begin{split}
    \LBCE^{(i)} =~& -\ytrue\log (\ypredi) \\
	& + \left(1-\ytrue\right)\log (1-\ypredi),
\end{split}
\end{equation}
with \ypredi (\ytrue) being the output (target) value of that classifier.
These BCE terms quantify the overall classification performance of the two classifiers.
Using the BCE form for the classifiers penalizes overly confident misclassifications.

The second term, \LDisCo, calculates the DisCo~\cite{Gabor:2007} value for the two classifier outputs.
This value is bounded between zero and one, with zero representing two independent discriminators and one representing two variables that are maximally correlated.
As described in \Sec{ABCD}, DisCo is more sensitive to nonlinear correlations than the standard Pearson correlation measure~\cite{Pearson}.
Thus, DisCo can be a more powerful measure for quantities like NN outputs that can exhibit nonlinear correlations.
A detailed description of the DisCo calculation is provided in Ref.~\cite{Kasieczka:2020yyl}.

Lastly, the \LClosure term quantifies the deviation from the assumed agreement of \NApred and \NAreg, \ie, the nonclosure of the ABCD method.
Including the nonclosure of the ABCD method in the loss function allows the network to explicitly optimize its performance for the ABCD method itself.
If the model is trained to classify events such that the ABCD method is satisfied by the classified populations of events, then by extension, the classifier outputs must be independent.
The nonclosure loss term is defined as:
\begin{equation}
\label{eq:closure}
    \LClosure = \left(\frac{\NAreg\NDreg-\NBreg\NCreg}{\NAreg\NDreg+\NBreg\NCreg}\right)^2.
\end{equation}
The numerator comes from \Equ{abcdRelation}, which implies $\NAreg\NDreg - \NBreg\NCreg = 0$.
The denominator $\NAreg\NDreg + \NBreg\NCreg$ serves to normalize the range of possible values with respect to other loss terms in case of nonclosure, whose magnitude depends on the $N$ yield values.
The expression is squared to ensure positive-definite values.
In terms of the explicit nonclosure definition in \Equ{nonclosure}, the nonclosure loss term can also be written as:
\begin{equation}
    \label{eq:nonclosureVsLoss}
    \LClosure = \left( \frac{\nonclosure}{2-\nonclosure} \right)^{2}.
\end{equation}
From this form of the nonclosure loss term, it can be seen that a nonclosure value of zero maps directly to a loss value of zero.
This term is computed only from background events, for which \Equ{abcdRelation} is intended to apply.

\Equation{closure} is entirely calculated from event counts, which are inherently discrete.
As such, \Equ{closure} can exhibit discontinuous and thus nondifferentiable behavior.
For example, it is possible that certain regions of the ABCD plane contain no events, which can cause undefined behavior due to division by zero.
For loss minimization via gradient descent, which requires the loss function to be differentiable, it is thus not possible to include \Equ{closure} directly in the loss function.
To address this problem, the method of counting the number of events in each ABCD region is slightly modified.
Mathematically, the counting of events can be visualized as applying a two-dimensional Heaviside step function $H$, whose edges correspond to the edge values defining the ABCD regions.
For region A, this calculation is:
\begin{equation}
\label{eq:heaviside}
    \NAreg = \sum_i^{\Ntot} H(\sonei-\bone) H(\stwoi-\btwo),
\end{equation}
where the sum is over all \Ntot events in the full plane.
Either an event is inside the boundaries of a region and counted (1) or it is outside and not counted (0).
Such a formulation makes clear that the discontinuous behavior lies in this binary counting choice.

Therefore, instead of using a step function, a two-dimensional sigmoid function can be used to smooth out this hard boundary.
This function takes the generic form:
\begin{equation}
\label{eq:sigmoid}
    \sigma(\sone,\stwo,\bone, \btwo) = \frac{1}{\left(1+\re^{-\aone(\sone-\bone)}\right)\left(1+\re^{-\atwo(\stwo-\btwo)}\right)},
\end{equation}
where the scale parameters \aonetwo control the sharpness of the transition from the minimum (0) to the maximum (1), the boundaries \bonetwo are the locations of the transition midpoint, and \sonetwo correspond to the values of the independent variables that define the ABCD plane.
The variables \aone and \atwo are, in general, separate hyperparameters to be optimized.
Recognizing the symmetry of the ABCD plane, \aone and \atwo are set here to the same value, $a$, for simplicity.
As a result, the calculation of the number of events in each region is transformed, following the form of the \Areg region from \Equ{heaviside}:
\begin{equation}
\label{eq:sigmoidA}
\begin{aligned}
    \NAreg &\to \sum_i^{\Ntot} \sigma(\sonei, \stwoi, \bone, \btwo), \\
	\NBreg &\to \sum_i^{\Ntot} \sigma(-\sonei, \stwoi, -\bone, \btwo), \\
	\NCreg &\to \sum_i^{\Ntot} \sigma(\sonei, -\stwoi, \bone, -\btwo), \\
	\NDreg &\to \sum_i^{\Ntot} \sigma(-\sonei, -\stwoi, -\bone, -\btwo).
\end{aligned}
\end{equation}

\begin{figure}[!htp]
   \centering
   \includegraphics[width=0.49\textwidth]{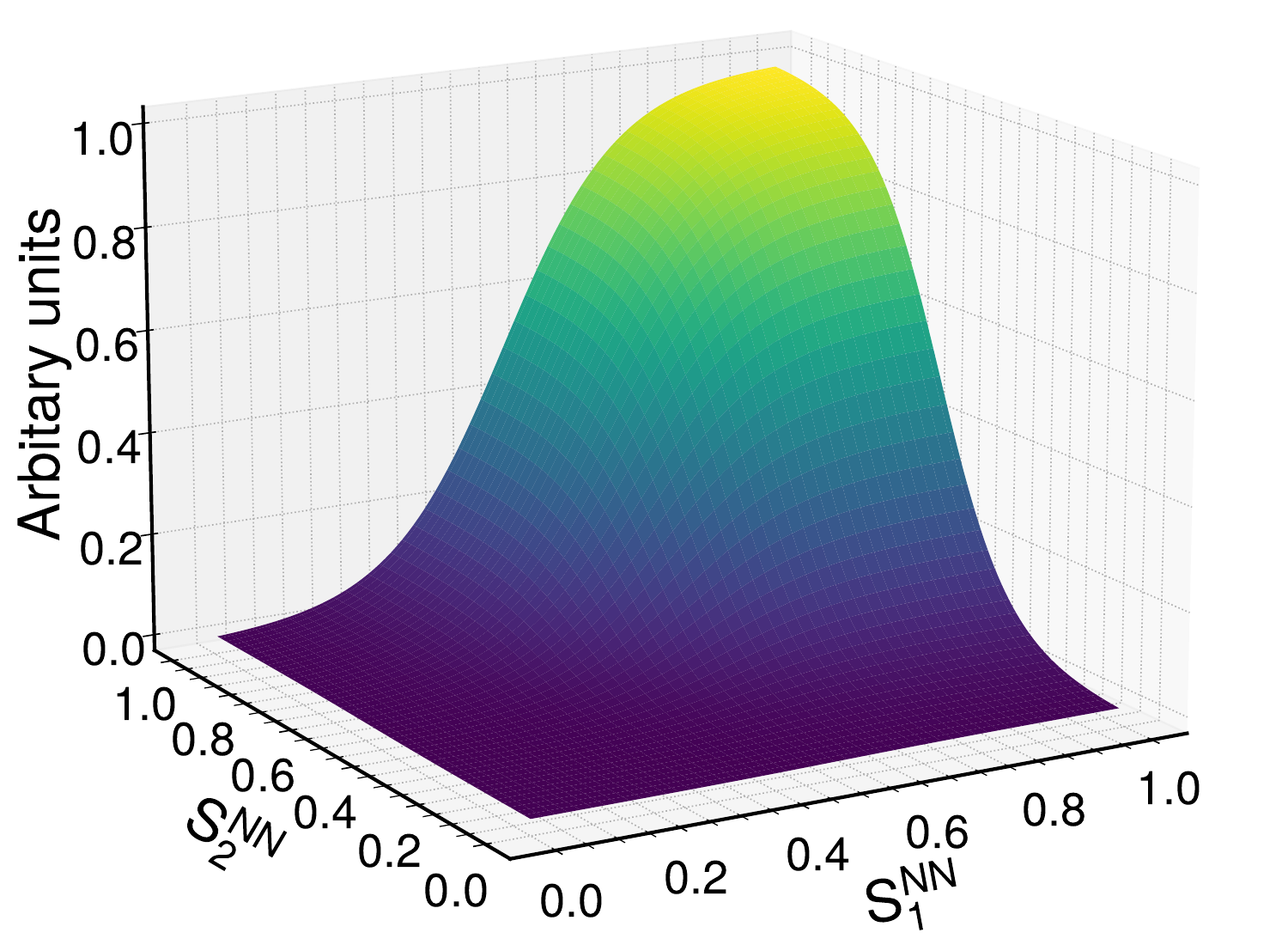}
   \includegraphics[width=0.49\textwidth]{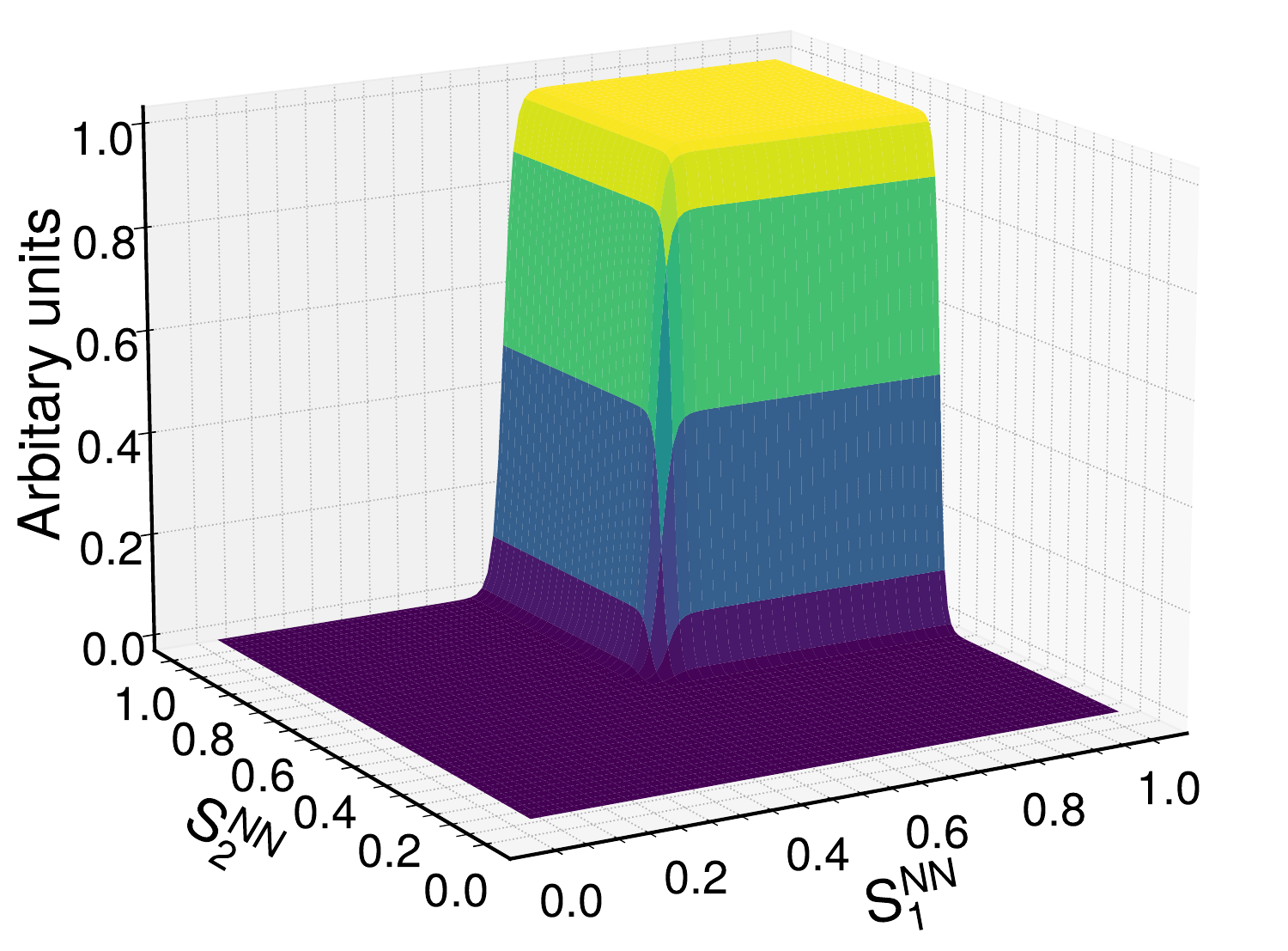}
   \includegraphics[width=0.49\textwidth]{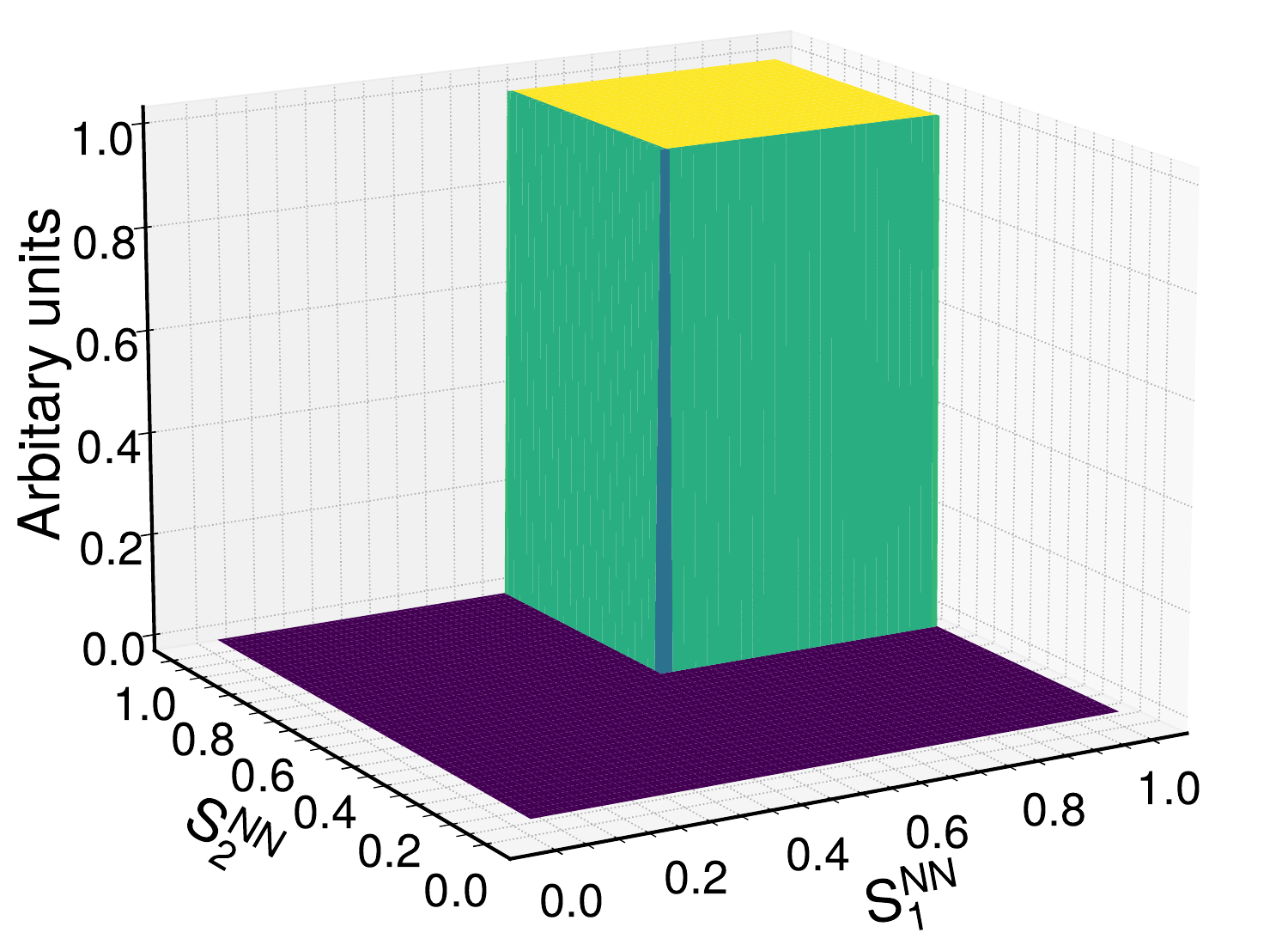}
   \caption{An example of the sigmoid function in \Equ{sigmoid} with choices for the scale parameters $a$ of 10 (\cmsUpperLeft), 100 (\cmsUpperRight), and $\lim a\to\infty$ (lower).
   The boundaries are $\bonetwo = 0.5$. The sigmoid distribution shown in the \cmsUpperRight is used for the ABCD region event counts in the following studies.
   \label{fig:sigmoids}}
\end{figure}

Using a two-dimensional sigmoid function (\Equ{sigmoid}) to weight events introduces continuity and differentiability to the nonclosure loss term (\Equ{closure}), ensuring the necessary behavior for use in gradient descent optimization.
With this sigmoid function, events can be slightly outside of the precise bounds of a region and still contribute to the region's event count with a small weight, strongly suppressed by the distance from the given region. 
Examples of how this two-dimensional sigmoid function appears for different choices of the scale parameter $a$ are shown in \Fig{sigmoids}.
The scale parameter value $a = 100$ was found to provide the best closure performance in general.
When optimizing $a$ for a specific application, both \LClosure and the nonclosure as a function of the choice of ABCD boundaries should be monitored.
If $a$ is too low, the minimization of \LClosure will be ineffective, because the amount of mixture between each region will be too large to achieve small nonclosure.
If $a$ is too high, the minimization of \LClosure will only be successful for the specific boundary values used during training, without sufficient generalization.
Choosing \bone and \btwo randomly for each batch during the training procedure, where a batch is a subset of the input data, is also important to improve the generalization of the closure.

With all of the loss terms in \Equ{totalLoss} defined, we provide a qualitative understanding of the interplay between the three terms to characterize the fundamental nature of this challenging optimization problem.
The \LBCE term pulls background toward the \Dreg region and pulls signal toward the \Areg region,
while \LDisCo and \LClosure shape the distributions to enforce the ABCD method requirement that the two discriminants be independent.
The \LBCE term should dominate the total loss, as it most directly relates to the performance of the network in separating signal from background.
However, the contributions of \LDisCo and \LClosure to the total are nonnegligible throughout the training, as they mold the discriminant shapes to ensure that robust background predictions can be performed.
Quantitative demonstrations of these principles are described in \Secs{traindetails}{closeanddisco}, using the example of a search for stealth supersymmetry (SUSY, \Sec{susyexample}).

\subsection{The modified differential method of multipliers}
\label{sec:mdmm}

Hyperparameter tuning is not guaranteed to lead to the desired optimum
in multiobjective optimization problems, such as minimizing the loss function in \Equ{totalLoss}.
More specifically, in multiobjective optimization, a solution, or point in objective space, is said to be Pareto-optimal if there is no direction in which one of the objectives can be improved without worsening at least one of the other objectives.
The set of all Pareto-optimal solutions achievable by a model is known as the Pareto front~\cite{Lotov2008}. 
The shape of this front is generally determined by the specific application and the NN model employed but is unknown a priori.
Thus, a search of some kind to find the desired optimum along this front is necessary. 

Using the typical approach of minimizing linear combinations of losses with gradient descent while tuning the hyperparameters $\lambda_{i}$, denoted the $\lambda$ method, does not necessarily guarantee access to the entire Pareto front. 
In general, convex Pareto fronts can be fully explored via the $\lambda$ method, while nonconvex Pareto fronts cannot.
In the case of a nonconvex Pareto front, the $\lambda$ method can only access the convex hull of the front, and under specific conditions, it is even possible that one of the losses is not minimized at all.
Even in the case of a convex Pareto front, finding an optimal training using the $\lambda$ method may require a large number of training iterations.
Also, as previously noted, the shape of the Pareto front for a given problem cannot be determined in advance.
Therefore, ensuring optimal results from a model with a multiobjective loss term may not be feasible when computational resources are limited. 

\begin{figure*}[hbtp]
    \centering
        \includegraphics[width=1.0\textwidth]{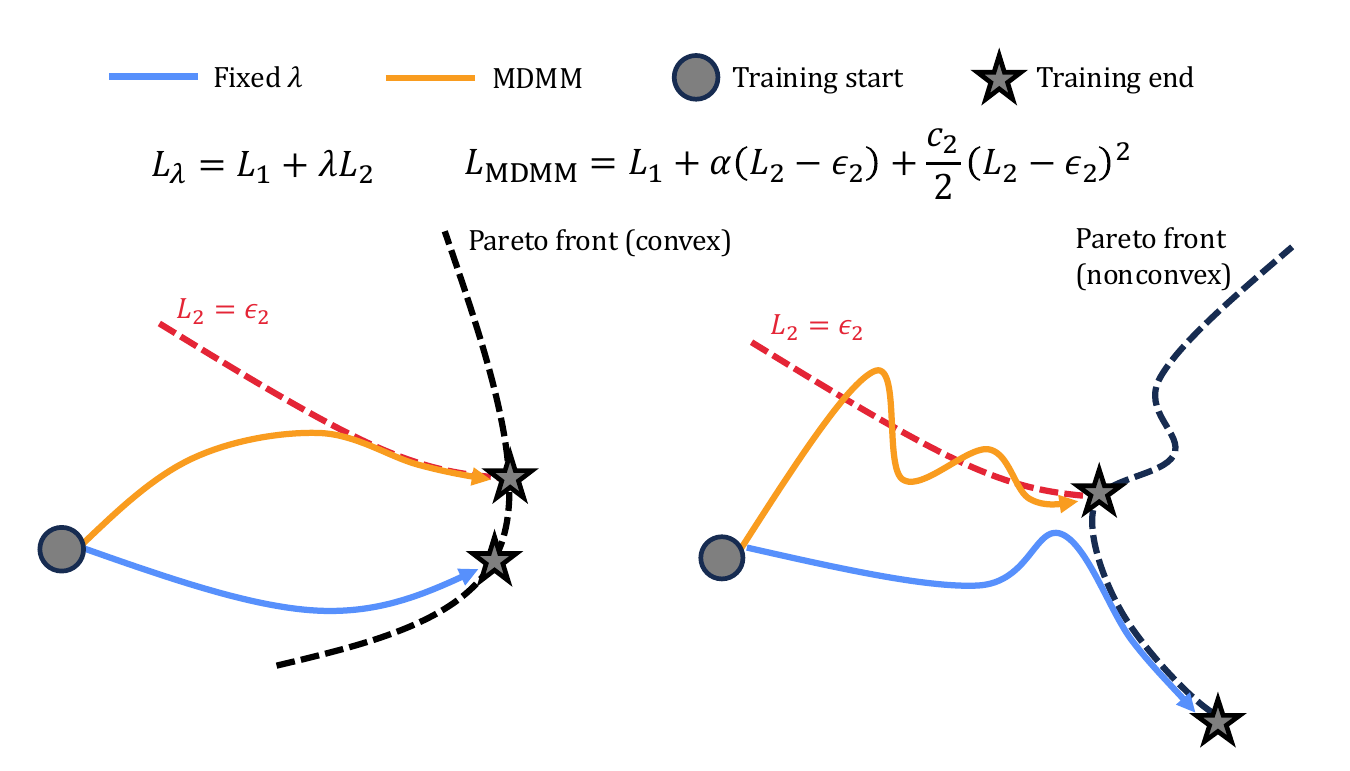}
    \caption{Schematic illustration of the training path of the NN in the space of the learned parameters for the $\lambda$ method (blue) and MDMM (orange). The case of an equality constraint on a generic loss component $L_2$ ($L_2 = \epsilon_2$) is shown. Both convex (left) and nonconvex (right) Pareto fronts, represented by dashed black lines, are shown. The dashed red line represents the subspace on which the condition $L_2 = \epsilon_2$ is met.
    \label{fig:cartoon_dmm_approaches}}
\end{figure*}

The MDMM~\cite{Platt:1987} approach offers the most robust solution for nonconvex Pareto fronts
and is particularly well suited for tackling the complex minimization problem described in this paper. 
This approach not only provides a means to explore the entire Pareto front, but also reduces the number of training iterations to find a usable solution. 

The core principle of the method is to reformulate the multiobjective optimization problem as a constrained optimization problem. 
One loss term is chosen as the function to minimize, while the other loss terms are treated as constraints via the Lagrange multiplier formulation. 
In the context of the ABCD method, the BCE loss is chosen as the function to minimize, while the DisCo and nonclosure losses are treated as constraints:
\begin{equation}
    \label{eq:constrained_optimization}
      \begin{split}
        \LClosure(\theta) & \leq \epsClosure, \\
        \LDisCo(\theta) & \leq \epsDisCo,
      \end{split}
\end{equation}
where $\theta$ represents the network weights and each $\epsilon$ is a hyperparameter representing the constraint value, replacing the corresponding $\lambda$ hyperparameter from the $\lambda$ method.
The Lagrangian function to minimize is then defined as:
\begin{equation}
    \label{eq:lagrangian}
    \begin{split}
        \mathcal{L}(\theta, \lambda) =~& \LBCE(\theta) \\
		& + \alpha_{\text{nonclosure}} \left(\LClosure(\theta) - \epsClosure\right) \\
        & + \alpha_{\text{DisCo}} (\LDisCo(\theta) - \epsDisCo),
    \end{split}
\end{equation}
where the Lagrange multipliers $\alpha$ are learned during the training.

The Lagrangian function is optimized with respect to the network weights and the Lagrange multipliers simultaneously.
After backpropagation when the learned parameters are updated, the network weights are updated in the direction opposite the gradient (gradient descent), while the Lagrange multipliers are updated toward the gradient (gradient ascent).
This is done because the optima for constrained minimization problems are typically saddle points in the gradient space; by inverting the update direction for the Lagrange multipliers, the optima become attractors for the differential equations that minimize \Equ{lagrangian}.

The final component of the MDMM formulation adds a quadratic penalty term to \Equ{lagrangian} for each constrained loss $i$: 
\begin{equation}
  \label{eq:quadraticConstraints}
  \frac{c_i}{2}(L_{i}(\theta) - \epsilon_{i})^{2}.
\end{equation}
This term introduces a further hyperparameter, $c_i$, for each constrained loss, called the damping factor.
In exchange, convergence on the submanifold that satisfies the constraints is now ensured, regardless of the shape of the Pareto front.
The training does not depend strongly on the value of this hyperparameter.
A pictorial representation of the effect of each component of the MDMM technique on the network behavior is shown in \Fig{cartoon_dmm_approaches}.
This illustration shows the dependence of the $\lambda$ method on the shape of the Pareto front, the attractive force supplied by MDMM, and the ability of the damping component to drive convergence.

The MDMM approach provides several advantages for NN training.
First, it enables exploration of the Pareto front in its entirety, not only its convex hull.
Second, it consistently converges to the chosen optimal solution in the case of nonconvex Pareto fronts, while the $\lambda$ approach does not.
Finally, when a loss term has a physical meaning, MDMM directly constrains the loss to a desired meaningful value using the $\epsilon$ hyperparameter. 
For example, in the case of $\LClosure$, setting \epsClosure to 0.1 corresponds to approximately a $10\%$ systematic uncertainty in the nonclosure, if no other correction is applied. 
By contrast, in the $\lambda$ method, the tuned hyperparameters only relate to the relative strength of each loss component, which may not have a clear physical meaning. 
Thus, MDMM not only allows, but encourages the definition of a ``target'' value for a given loss component, removing the need to scan in an additional dimension of hyperparameter space. 
In \Sec{mdmm_app_res}, the MDMM approach is applied to the \discotec method using the stealth SUSY data set, showing an improvement in the stability and robustness of the method.

\section{Case study: a search for stealth supersymmetry}
\label{sec:susyexample}

We demonstrate the benefits of the \discotec method in a search for new physics beyond the standard model (SM) of particle physics using proton-proton collisions at a center-of-mass energy of 13\TeV. 
Specifically, we search for the pair production of top squarks (\PSQt), the supersymmetric partner of the top quark, in the context of stealth SUSY models~\cite{Fan:2011yu, Fan:2012jf, Fan:2015mxp}. 
In the model we consider, the decay of the two top squarks results in two top quarks accompanied by six gluons.
An initial search for this signature was published by the CMS Collaboration in Ref.~\cite{CMS:2021knz}.
The collision events selected in this search contain a charged lepton from a top quark decay and at least seven jets, which are collimated sprays of particles originating from the hadronization of quarks or gluons produced in the interaction.
This final-state signature is quite challenging to distinguish from SM background processes, in particular the background from top quark pair production in association with additional jets (\ttjets). 
This similarity between signal and background restricts the sensitivity of simple selections applied to individual high-level variables.

An NN can much more efficiently determine and take advantage of subtle differences between signal and background, as was demonstrated for this type of model in Ref.~\cite{CMS:2021knz}.
However, the search was limited by systematic uncertainties related to the modeling of the \ttjets background in simulation, resulting in an inconclusive statement about a deviation between the predicted background and the observed data.
Accordingly, an NN using the \discotec approach is a particularly attractive solution to achieve strong discrimination between signal and background while improving the robustness of the background estimation by increasing the reliance on observed data rather than simulated events.
The application of the \discotec method to the CMS search for stealth SUSY is showcased in Ref.~\cite{SUS-23-001} with significantly improved performance compared to Ref.~\cite{CMS:2021knz}.
Here, we will use the stealth SUSY signature to characterize features of the \discotec method and to illustrate how the method can be validated for use with real collision data.

\section{Implementation}
\label{sec:implementation}

\subsection{Model training}
\label{sec:traindetails}

For the case study, the \discotec model was extended with a regression component that precedes the dual binary classifiers. 
This regression aims to predict the mass of the top quark (for background) or top squark (for signal) in an event, and was found to improve the classification performance of the network for this application.
Specifically, adding the mass regression allows for better classification of events with top squarks at low masses ($\mstop < 600\GeV$), which are kinematically similar to background events.
The inputs that are provided to the mass regression component are also passed (along with the mass regression output itself) to each binary classifier during training and inference.
The total loss function thus includes an additional term, \LMR, which takes the form of the squared error between the predicted particle mass, \mpred, and the true mass, \mtrue:
\begin{equation}
\label{eq:massRegLoss}
    \LMR = \frac{(\mtrue - \mpred)^{2}}{\GeVns^2}.
\end{equation}
This additional loss term is not strictly required to implement the \discotec framework.
For the purposes of demonstrating the validity of the \discotec method, the output of the mass regression can simply be considered as another model input.
All components of the neural network are jointly optimized during training by minimizing the total loss function.

Two versions of the model are trained.
The first model was trained in the \textsc{Keras}~\cite{keras} software framework (v2.4.3) with the \textsc{TensorFlow}~\cite{Abadi:2016kxi} Python package (v2.2.0) used to construct the model and manage the training environment.
The second model was constructed and trained using the \PYTORCH~\cite{Paszke:2019} Python package (v2.4.1). 
The latter framework was adopted for ease of implementation of MDMM.
Both models show similar performance in terms of signal vs. background discrimination as well as nonclosure. 
The training information below applies to both of the implementations. 
The gradient-based optimizer, \ADAM~\cite{Kingma:2014vow}, is used during both trainings for loss minimization.

The discriminant component of the network is composed of two fully connected NNs with one hidden layer each.
An additional hidden layer is placed before the discriminant component to improve the decorrelation behavior of the network. 
Each hidden layer is followed by both a batch normalization and a dropout layer.
The dropout layers randomly remove 30\% of the nodes in the hidden layer to prevent overtraining. 
There are 70 input nodes, 200 nodes per hidden layer, and two output nodes (one for each discriminant).
The inclusion of additional hidden layers was not found to provide significant improvements in performance.

The input features used in training the \discotec network to distinguish the stealth SUSY signal from SM backgrounds include both low- and high-level physical quantities. 
The low-level features include the four-momenta of jets and the type of particle from which each jet likely originated.
The high-level features are based on the relative orientation and momenta of the jets, in order to give a global view of the energy flow in an event.
In total, each training event is represented by 70 features.
The events included in the training must satisfy a baseline set of relatively loose requirements that are designed to eliminate obvious sources of background.

For the primary \ttjets background, events are generated with the next-to-leading order (NLO) \POWHEG v2.0 generator~\cite{powheg1,powheg2,powheg3,powhegttbar} and normalized using production cross sections calculated to approximate next-to-NLO (NNLO) accuracy~\cite{Czakon:2011xx}. 
Additional samples are simulated using the same generator with several physics model parameters varied such as those that govern the evolution of particle showers in the collision.
For the signal, top squarks are produced with \MGvATNLO v2.6.5~\cite{madgraph} at leading order with up to two additional partons at the matrix element level.
The signal production cross sections are calculated at approximate NNLO plus next-to-next-to-leading-logarithmic accuracy~\cite{signalxsecNNLO,signalxsecNNLL}.
The decays of the top squarks are handled by \PYTHIA v8.240~\cite{pythia}.
Parton distribution functions computed at NNLO accuracy from NNPDF3.1~\cite{NNPDF} are used for all processes.
For both \ttjets and signal, \PYTHIA is used for parton showering and hadronization, with the underlying-event tune CP5~\cite{Sirunyan_2020}. 
The CMS detector response is simulated using \GEANTfour~\cite{GEANT}. 

For signal, events are simulated with a range of top squark masses from 300 to 1400\GeV in increments of 50\GeV.
For both background and signal events, additional variations from the uncertainties in the jet energy scale and jet energy resolution are considered.
Of the total training set, 80\% of events are used for training while the remaining 20\% are split evenly between test and validation sets.
Overall, there are ${\approx}100$ million background events and ${\approx}1$ million signal events available in the training set.

For each background and signal sample, events are divided roughly equally among five categories corresponding to the number of jets (\njets) in the event.
The \njets value is the primary discriminant between signal and background, where signal on average has a higher \njets.
Thus, sensitivity to signal is maximized for events in the highest \njets category.
However, there are more events with lower \njets relative to the number of events with higher \njets, so each \njets category is equally sampled to ensure that the network's performance is homogenized across \njets.

The training events are split into batches, maintaining the equal division among the \njets categories.
The batches are also evenly split between background and signal events, each of which contains a mixture of events generated with the nominal simulation settings and events generated with modeling variations.
The DisCo and nonclosure losses are both computed on a per-batch basis; random boundaries are used for the nonclosure. 
Choosing an appropriate batch size amounts to balancing between poor nonclosure estimates for small batch sizes and high GPU memory requirements for large ones.
Likewise, the optimal learning rate is also affected by the interplay of the loss terms.
An overly large learning rate inhibits the network's ability to achieve minimal nonclosure across the plane, but one that is too small results in poor discriminating power or onerous training times. 
For the stealth SUSY training, a batch size of 4096 and a learning rate around $10^{-4}$ to $10^{-5}$ with the \ADAM optimizer are found to be optimal. 
However, care must be taken to monitor and tune these hyperparameters for any individual training, as they are application and optimizer dependent. 
The $\lambda$ scaling hyperparameters for each loss term are tuned on the training data set such that both the classification and nonclosure performance are reasonable for the analysis.

\begin{figure}[hbtp]
    \centering
        \includegraphics[width=0.49\textwidth]{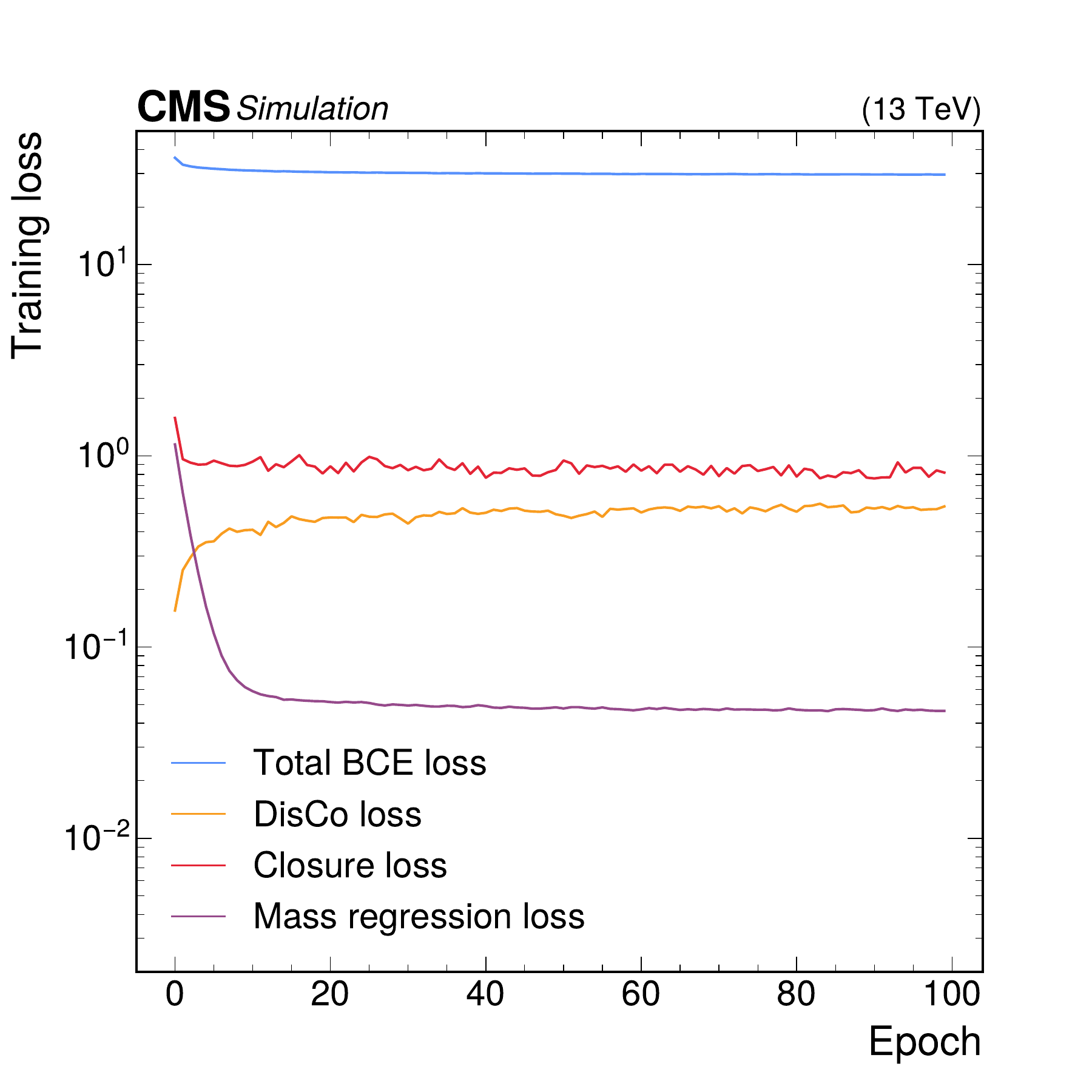}
        \includegraphics[width=0.49\textwidth]{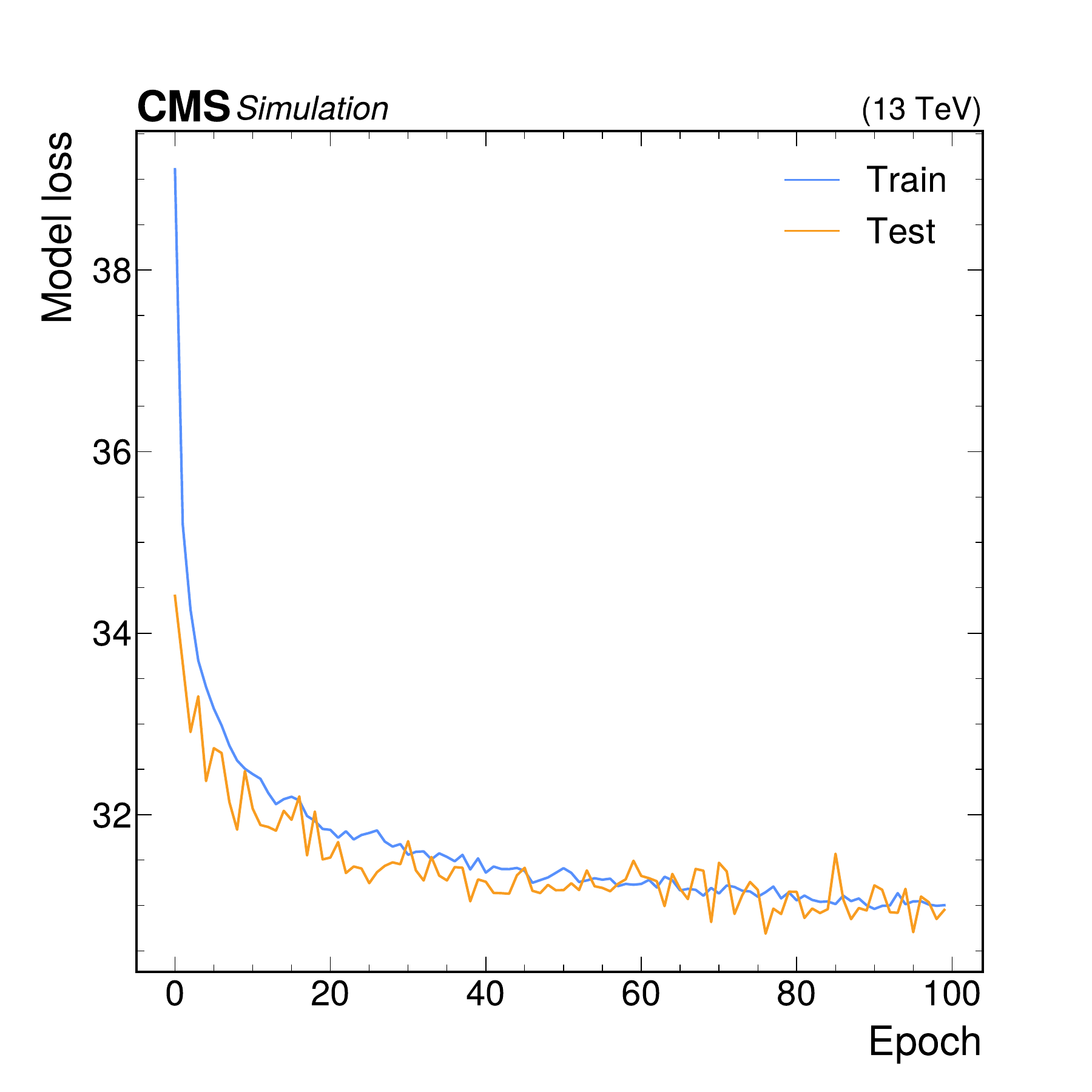}
    \caption{The values of the individual loss components scaled by the corresponding $\lambda$ value computed on the training set (\cmsLeft) and the total loss computed on both the training and test sets (\cmsRight) as a function of epoch for the stealth SUSY case.
The network training is stopped after 100 epochs, after which neither the training nor test sample losses decrease.
The differences in magnitude between the different loss components result, in part, from the magnitude of their respective hyperparameters; the total loss smoothly decreases.}
    \label{fig:loss}
\end{figure}

\begin{figure}[htp]
    \centering
        \includegraphics[width=0.49\textwidth]{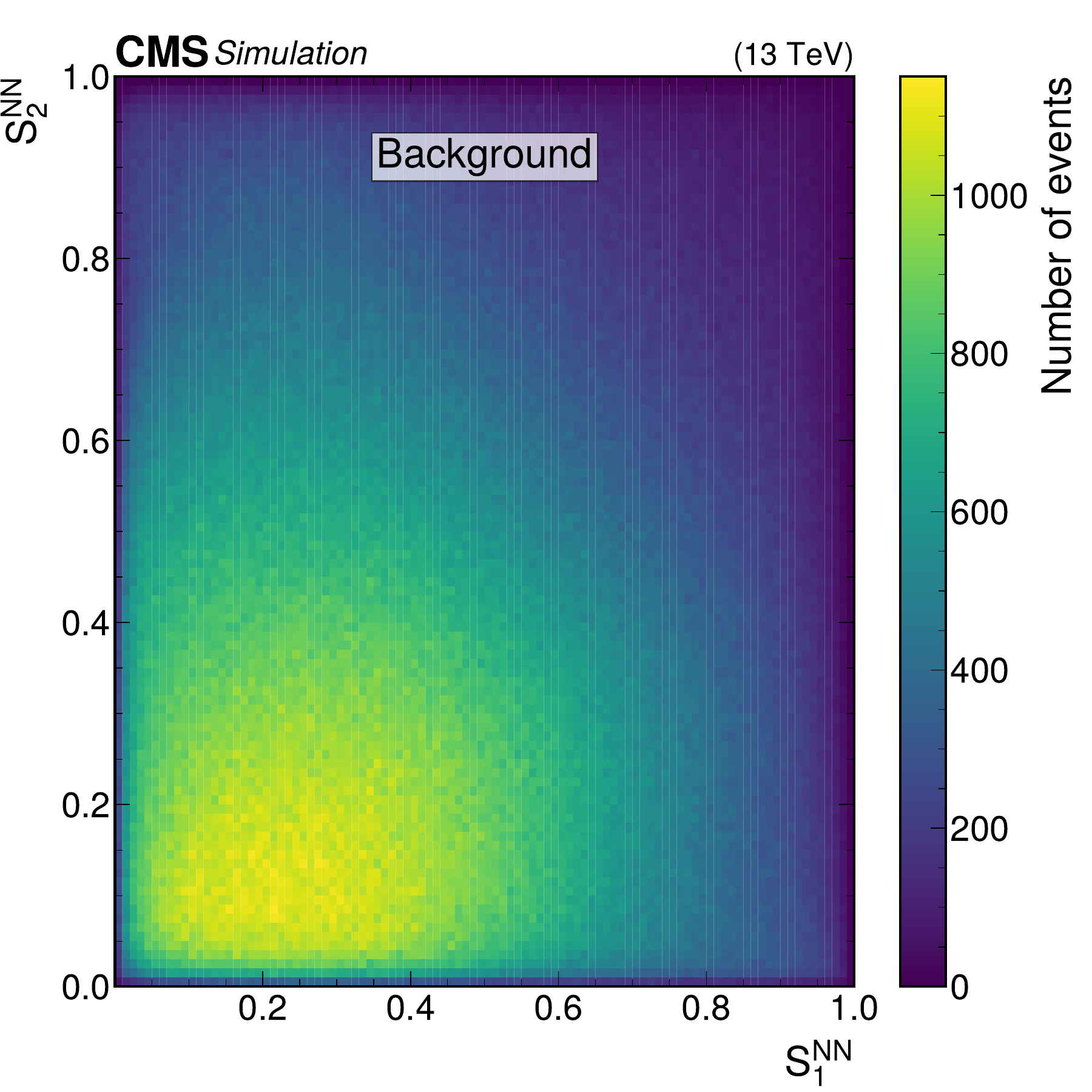}
        \includegraphics[width=0.49\textwidth]{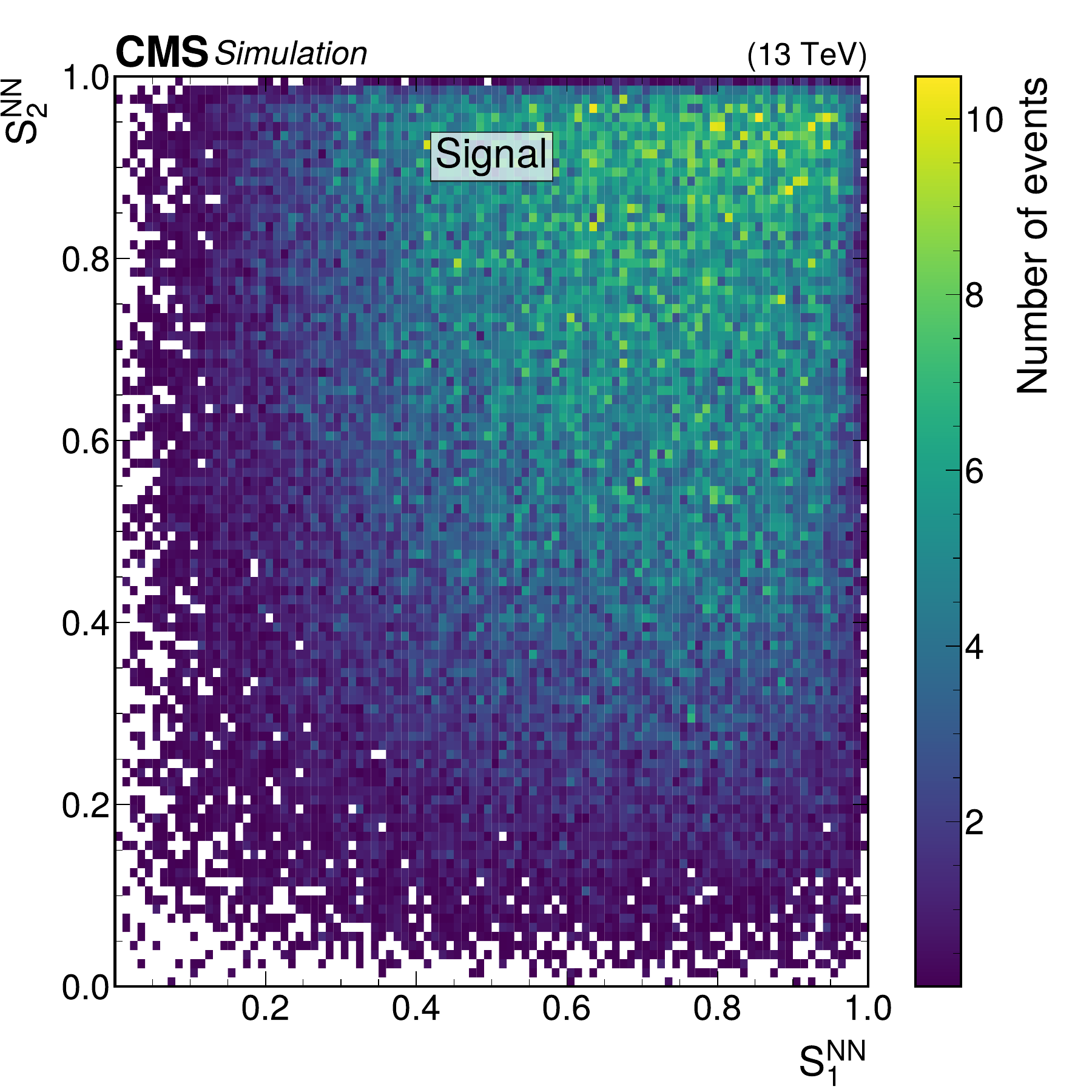}
    \caption{The distributions of background (\cmsLeft) and signal (\cmsRight) events in the two-dimensional discriminant plane for the final training of the \discotec model with the stealth SUSY training set.
    White bins contain no events.}
    \label{fig:bkgSig2D}
\end{figure}

\begin{figure}[htp]
    \centering
        \includegraphics[width=0.49\textwidth]{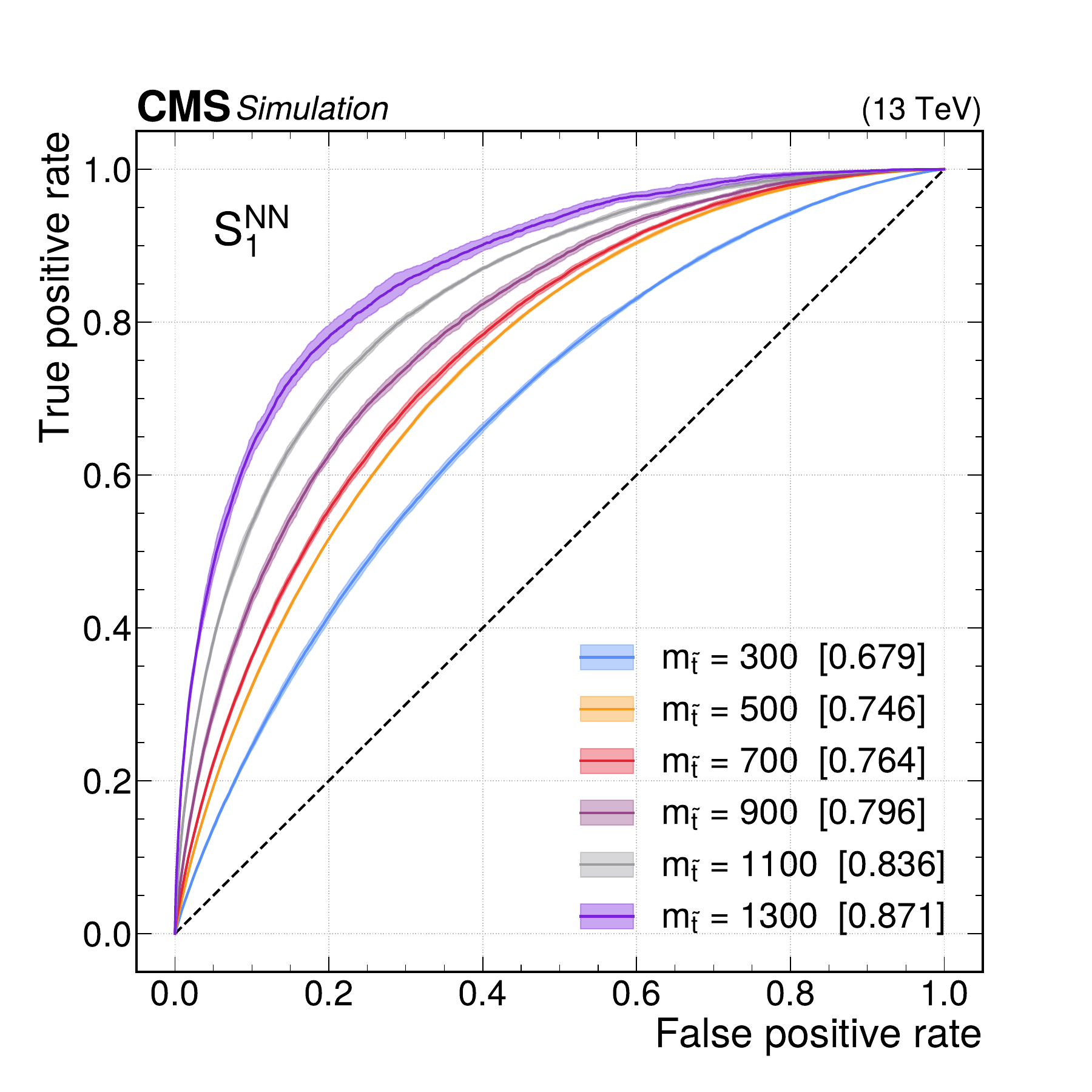}
        \includegraphics[width=0.49\textwidth]{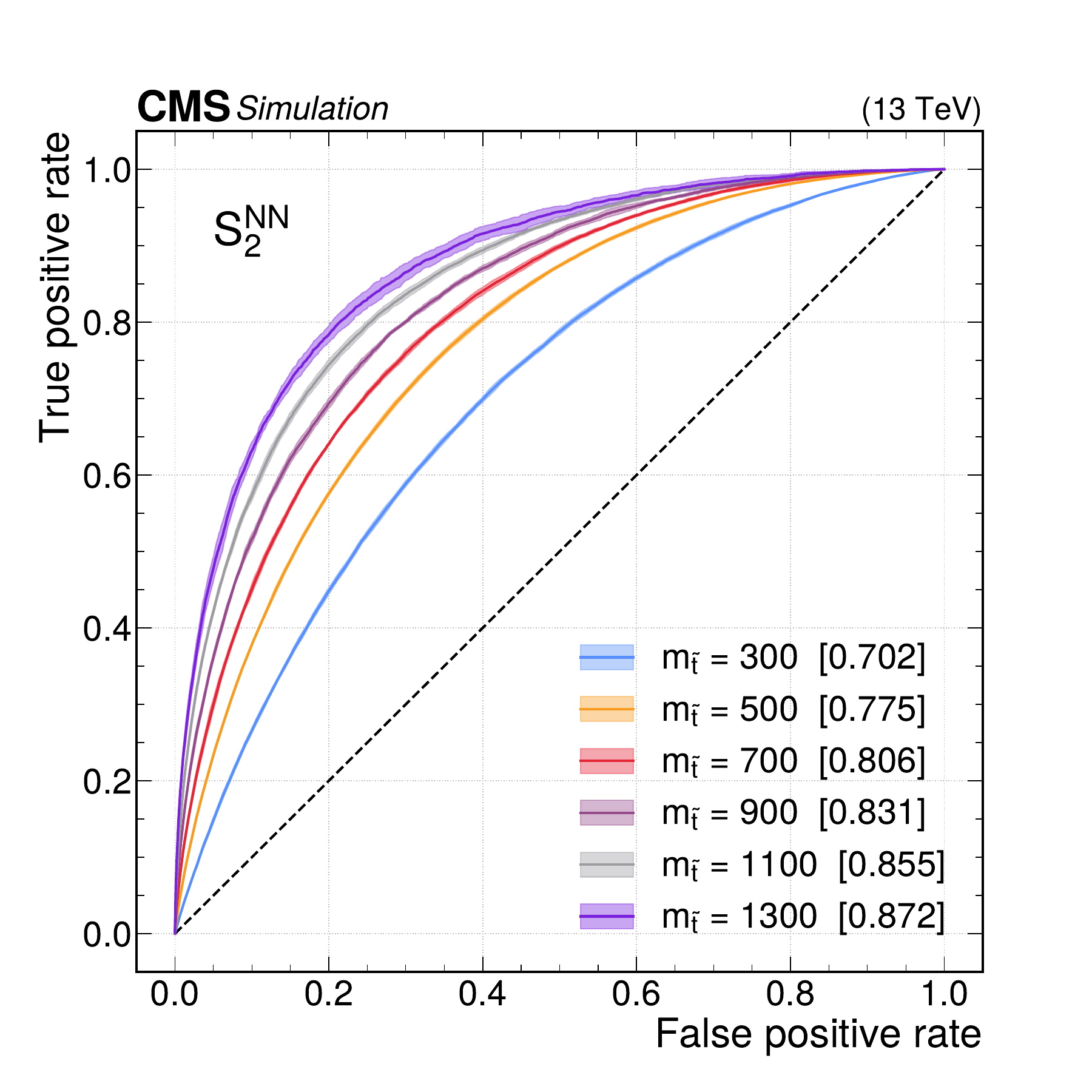}
    \caption{The ROC curves for \snni{1} (\cmsLeft) and \snni{2} (\cmsRight) from the \discotec model using the stealth SUSY training set, for different values of the top squark mass in units of {\GeVns}.
    The performance is measured using an eight-fold cross-validation;
    the shaded regions represent one standard deviation of the average performance of the true positive rate for each false positive value in the test.
    The black dashed line represents the ROC curve for random classification. 
	The values in brackets are the areas under the ROC curves.}
    \label{fig:ROCpermass}
\end{figure}

The loss per epoch during the model training is shown in \Fig{loss}.
No overtraining is observed, as the test sample loss closely matches the training loss.
While all loss components are intended to be minimized, the distance correlation loss increases as a function of epoch because improving binary classification necessarily results in higher correlation between the two discriminants.
The performance of the final training of the \discotec model is demonstrated in Figs.~\ref{fig:bkgSig2D} and~\ref{fig:ROCpermass}. 
The overall distribution of background and signal events in the two-dimensional discriminant plane in \Fig{bkgSig2D} has the desired qualities outlined in the previous sections.
The background shape peaks in the lower-left corner and shows strong decorrelation between the two discriminants.
Likewise, signal peaks in the upper-right corner and is separated from background.
\Figur{ROCpermass} shows the receiver-operator characteristic (ROC) curves for the two independent discriminants in terms of the true positive rate---classifying signal events as signal---and the false positive rate---classifying background events as signal.
There is strong discrimination between signal and background across the different signal mass points used in the training.
The best performance is achieved for the highest-mass signal events, since these are physically the most distinct from the background.
However, the addition of the mass regression is responsible for an increase in the area under the ROC curve of around 0.1 for both discriminants for low $\mstop$.

\begin{figure}[htp]
    \centering
        \includegraphics[width=0.49\textwidth]{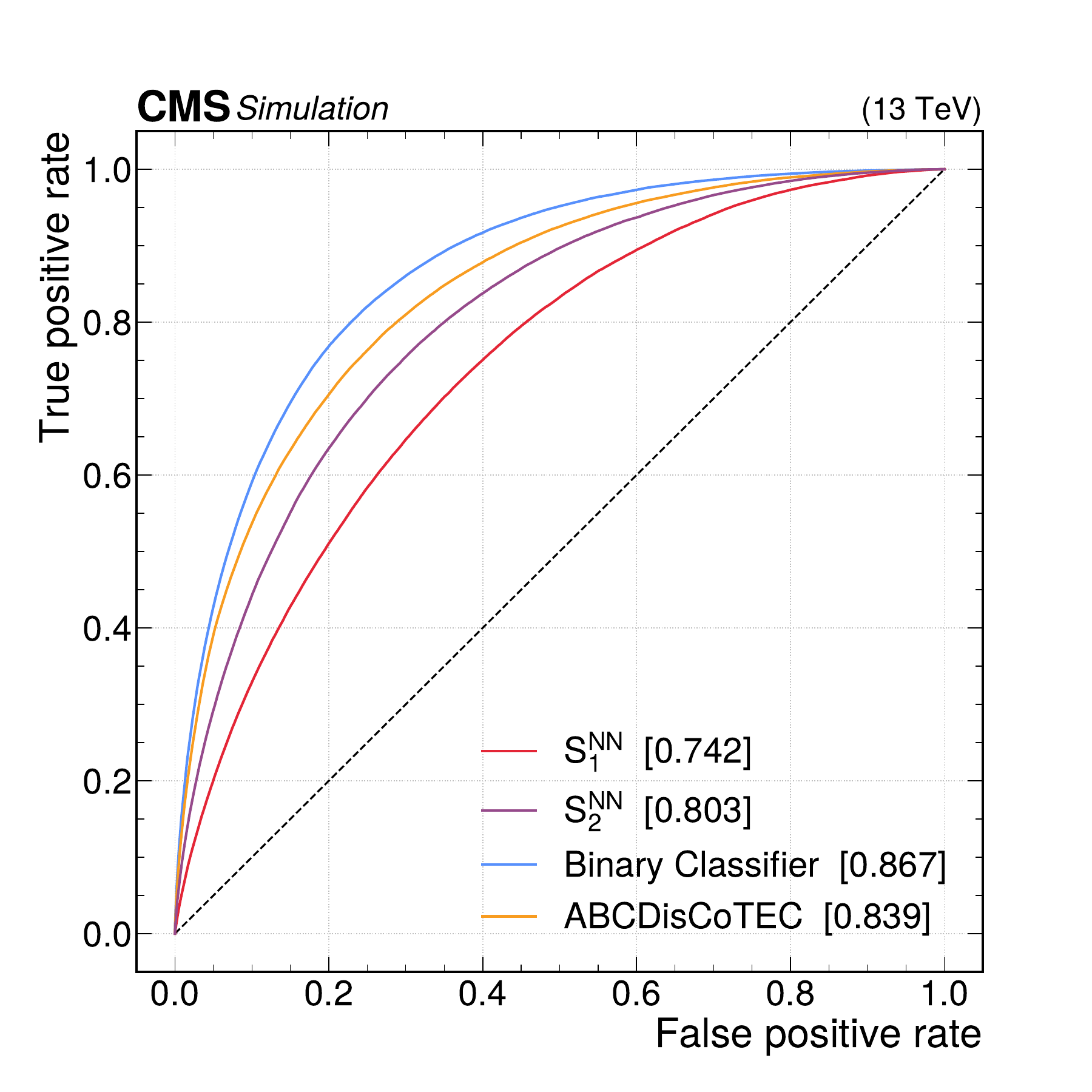}
    \caption{Comparison of ROC curves from a single binary classifier DNN, each discriminant from the \discotec model, and the two \discotec discriminants, for all top squark masses combined.
    The last entry is computed arbitrarily as the distance of the two discriminants from $(0,0)$.
    The black dashed line represents the ROC curve for random classification.}
    \label{fig:ROCSingleVsDouble}
\end{figure}

A performance comparison between a standard binary classification NN and the \discotec model is shown in \Fig{ROCSingleVsDouble}.
The results for the binary classifier are obtained by training the \discotec network with both \lamClosure and \lamDisCo set to zero.
This is equivalent to a single binary classification network, as the two outputs are no longer subject to decorrelation constraints. 
Along with this curve, three other ROC curves show the performance of the \discotec network: the two ROC curves for the two independent binary classifiers and one for an arbitrary combination of the two discriminants. 
This combination is calculated using a normalized distance from $(\snni{1}, \snni{2}) = (0, 0)$.
That is, the ROC curve for the combination of the two discriminants is calculated using thresholds placed on $\sqrt{\smash[b]{(\snni{1})^2 + (\snni{2})^2}} / \sqrt{2}$.
These results show that the impact of the extra constraints in the \discotec method on the classification performance is relatively small compared to the maximum achievable discrimination from a single binary classifier.

\subsection{Characterization of the model behavior}
\label{sec:closeanddisco}

Previous work~\cite{Kasieczka:2020pil} has shown that a model with only classification and DisCo loss terms could result in a background distribution meeting the validity requirements for the ABCD method: the two variables should be uncorrelated and the background events should be smoothly falling and localized at low values of each variable.
However, a model trained with just those two loss terms will not necessarily achieve adequate performance.
Undesirable solutions that violate the validity requirements can arise from two-dimensional distributions that have low distance correlation values but are not desirable for the background estimation procedure.
In this situation, finding a training result that avoids these pitfalls requires an exhaustive scan over a large portion of hyperparameter space.
The addition of the nonclosure loss term eliminates the need for an exhaustive scan by preventing undesirable solutions, thereby improving the overall training efficiency and correspondingly the efficacy of the background estimation method.
This section outlines the dynamics of training in the optimization space defined by \Equ{totalLoss} and how the various terms interact to achieve superior performance.

To characterize the synergy between the DisCo and nonclosure loss terms, a series of networks are trained using three separate total loss functions: one using only \LDisCo, one using only \LClosure, and one using both. 
All three total loss functions also include the BCE term for classification.
The scaling hyperparameters are scanned within a representative range of the best values for each of the configurations.
For the trainings using only \LClosure, the scanned values 
range from $\lamClosure = 1$ to 1000.
Values of \lamDisCo
from 5000 to 50,000 are selected for trainings using only \LDisCo.
For the combined trainings, the scanned ranges are
$\lamClosure \in [10,100]$ and $\lamDisCo \in [10,100]$.
All other hyperparameters are set to optimal values for each training configuration, defined as maximizing discrimination with $\nonclosure < 0.1$.

For the trainings using only \LDisCo, a significant correlation has been observed between the scaling hyperparameter value, learning rate, and batch size.
For this case only, the values of the batch size and learning rate are also scanned within appropriate ranges to ensure optimal performance and reduce overtraining. 
Batch sizes of 10,000, 16,000, and 20,000 and learning rates of $10^{-4}$, $10^{-5}$, and $10^{-6}$ are scanned for each of the \lamDisCo values.
The ranges for the scaling hyperparameters are intentionally chosen to be wide to include values that may not lead to ideal solutions.
This demonstrates the sensitivity of each loss configuration to changes in the values of the scaling hyperparameters. 

\begin{figure*}
  \centering
    \includegraphics[width=0.6\textwidth]{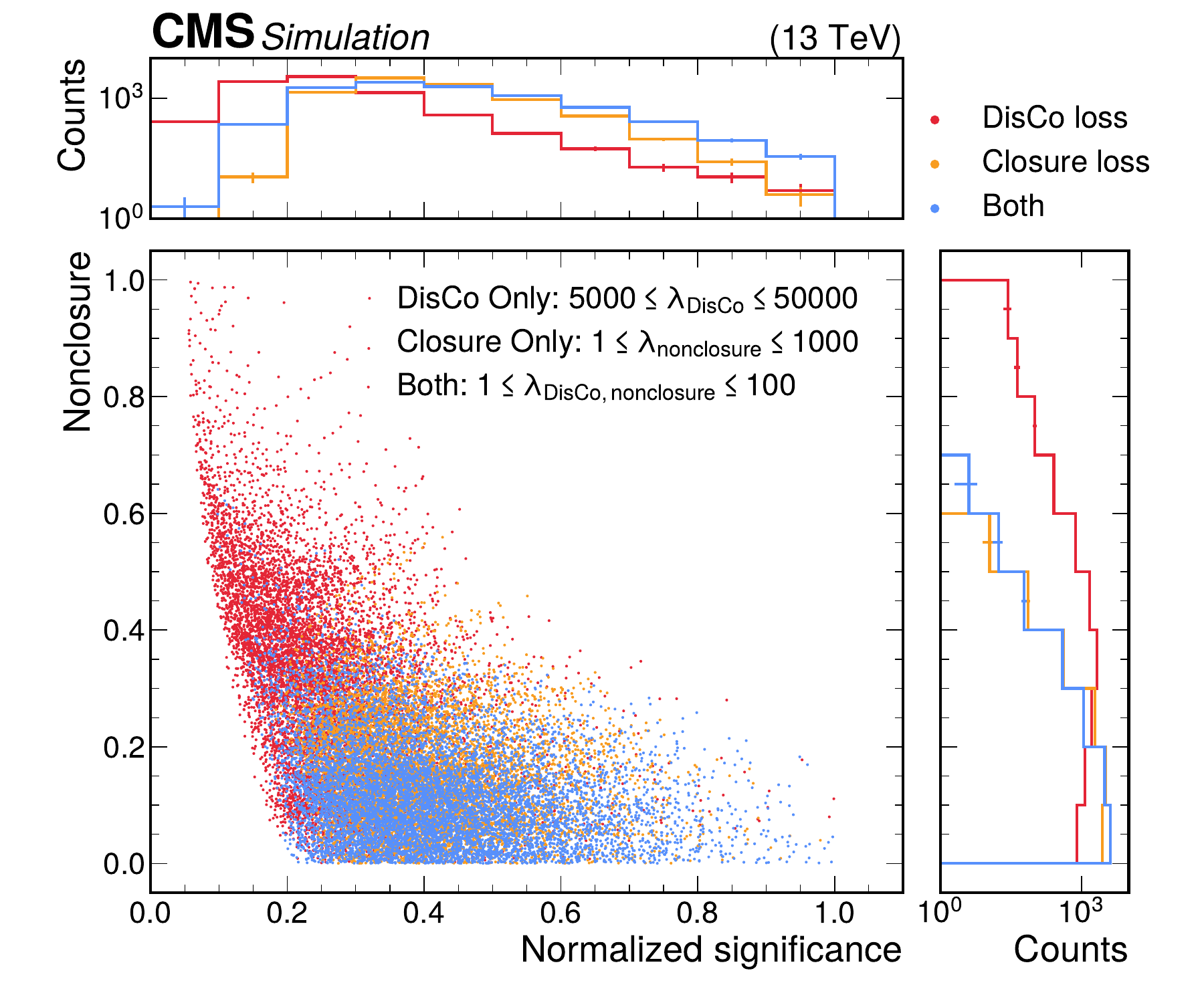}
  \caption{
Comparison of the nonclosure and normalized significance from the three loss function variations (DisCo only, nonclosure only, and DisCo and nonclosure)
for the stealth SUSY training set using a scan of boundaries over multiple trainings.
The significance values from \Equ{sigWithNonClosure} are normalized by the maximum significance value of the three training configurations.
}
  \label{fig:nonclosureSign}
\end{figure*}

The performance of the network is quantified for a grid of ABCD boundary choices.
These boundaries range from discriminant values of 0.2 to 0.8 in steps of 0.05 for both discriminants.
In addition to examining the nonclosure directly, the effect of the classification performance in separating signal from background is incorporated in an additional metric, the signal significance.
Here, the significance is computed as:
\begin{equation}
    \text{Significance} = \frac{\Nsig}{\sqrt{\Nbkg (1 + \nonclosure^{2})}}
    \label{eq:sigWithNonClosure}
\end{equation}
where \Nsig and \Nbkg are the numbers of simulated signal and background events, respectively, in region \Areg, and \nonclosure is the nonclosure for given boundaries.
Here, the nonclosure is taken directly as a relative systematic uncertainty in the estimated background yield, as it represents the deviation of the prediction from the correct value, and is therefore included as a quadratic term.
This significance metric, therefore, represents how a systematic uncertainty in nonclosure would affect the overall sensitivity to signal.
\Figur{nonclosureSign} shows these metrics for the three training configurations for each boundary in the grid.

We find that the loss function Eq.~\eqref{eq:totalLoss} results in both the lowest average nonclosure and the highest significance of the three loss functions.
It also shows greater stability in the closure performance than the other two variations, with few instances of nonclosure above 0.5.
The nonclosure-only loss function results in similar significance values for some boundary choices; however, this comes at the expense of higher average nonclosure, which affects the final sensitivity of the analysis.
In turn, for any given set of boundaries, the DisCo loss function has a higher concentration of results with high nonclosure and low significance.
The combined loss function shows improved robustness to the choice of boundary values, which reduces systematic uncertainties associated with nonclosure.
The combined loss function is thus more likely to produce a viable final training than either of the two components alone.

\subsection{Avoidable training failure modes}
\label{sec:failuremodes}

The main challenge in training an NN with a multiobjective loss function is determining the optimal configuration of hyperparameters to produce a network that meets all of the criteria for a given application.
While training the \discotec network, a number of distinct failure modes appeared that provide some additional insight into the behavior of the model and can offer guidance for understanding the nature of the hyperparameter space. 
This section outlines a selection of the most common failure modes and suggests how they can be avoided.

\begin{figure}[!htp]
    \centering
        \includegraphics[width=0.49\textwidth]{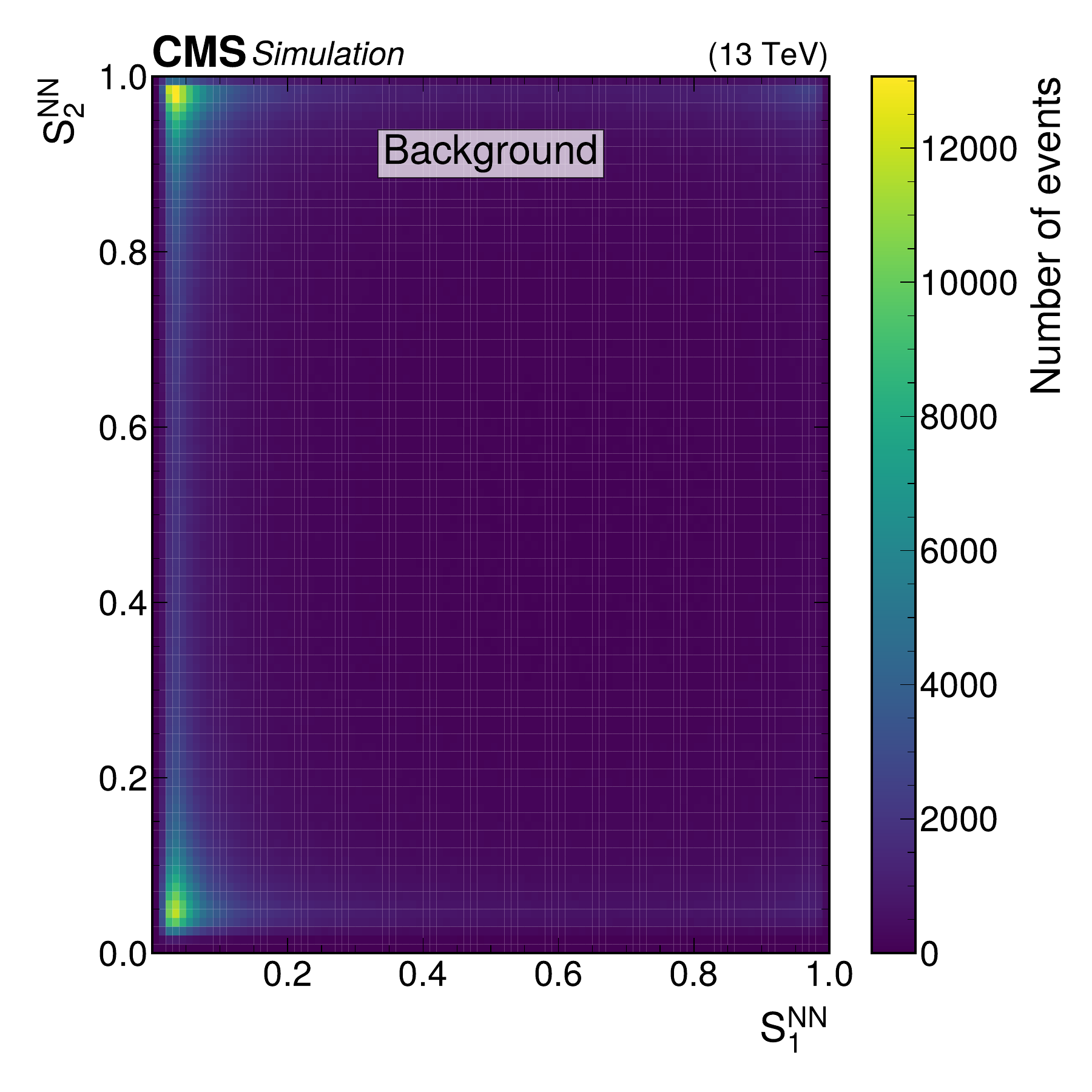}
        \includegraphics[width=0.49\textwidth]{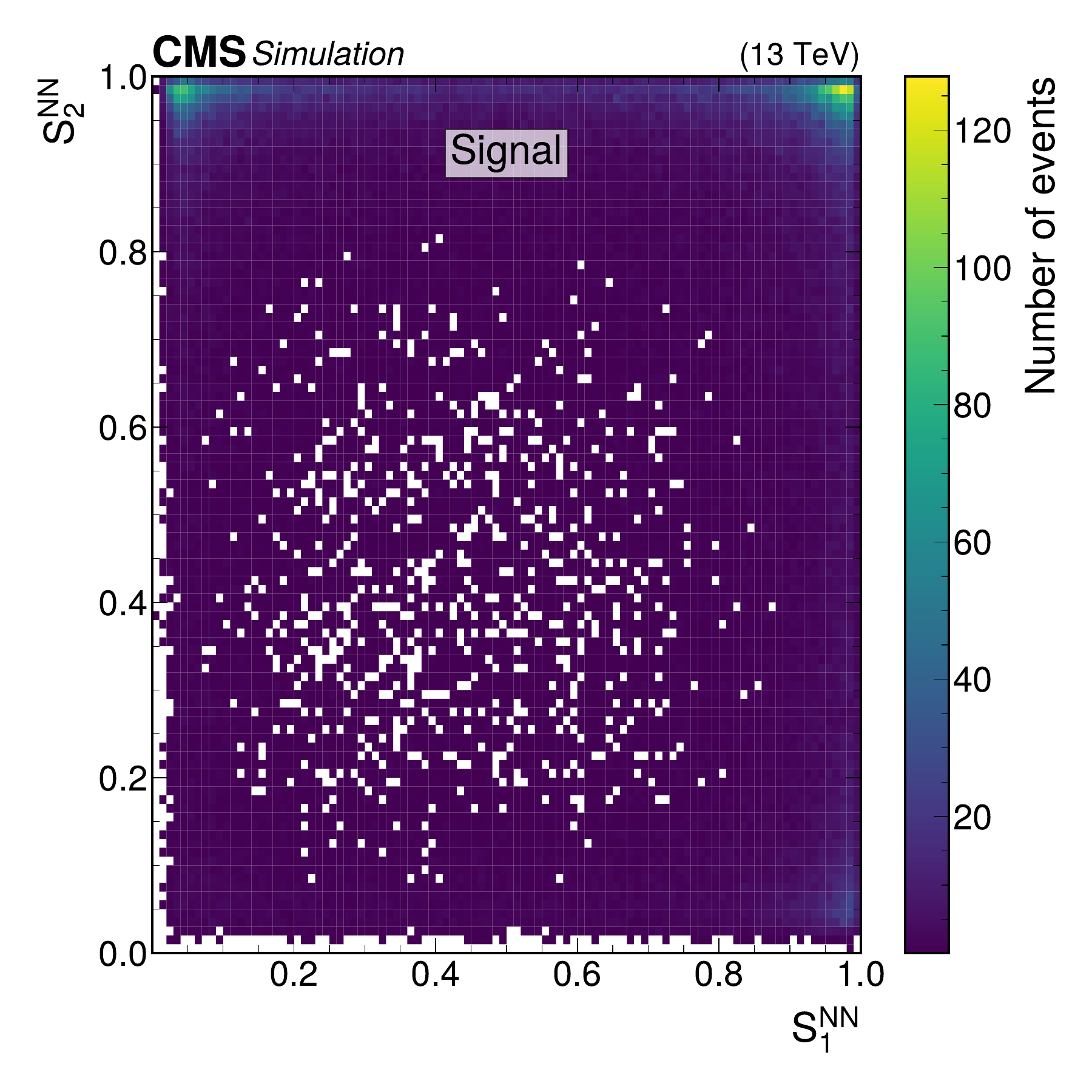}
    \caption{Examples of the ``four corners'' failure mode: background (\cmsLeft) and signal (\cmsRight) distributions in which the distance correlation is approximately zero. 
    A relatively large \lamDisCo value may result in convergence to this failure mode.
    White bins contain no events.}
    \label{fig:fourCorners}
\end{figure}

The first failure mode exhibits a clustering of discriminant values in some or all of the four corners of the ABCD plane, as shown in \Fig{fourCorners}.
Since the DisCo value is near zero and the ABCD relation is satisfied (nonclosure is zero), this distribution can arise when the hyperparameters associated with these two loss terms are too large.
Thus, DisCo and nonclosure alone are not sufficient to achieve meaningful performance, and the relative importance of the BCE term must be increased.
Specifically, for the stealth SUSY training set, it is observed that this effect occurs when the nonclosure or DisCo losses are larger than 10\% of the BCE component.
These solutions can be avoided by reducing \lamClosure and \lamDisCo while monitoring the values for each loss component individually as the training proceeds, to ensure that \LBCE is being appropriately taken into account.

\begin{figure}[htp!]
    \centering
        \includegraphics[width=0.49\textwidth]{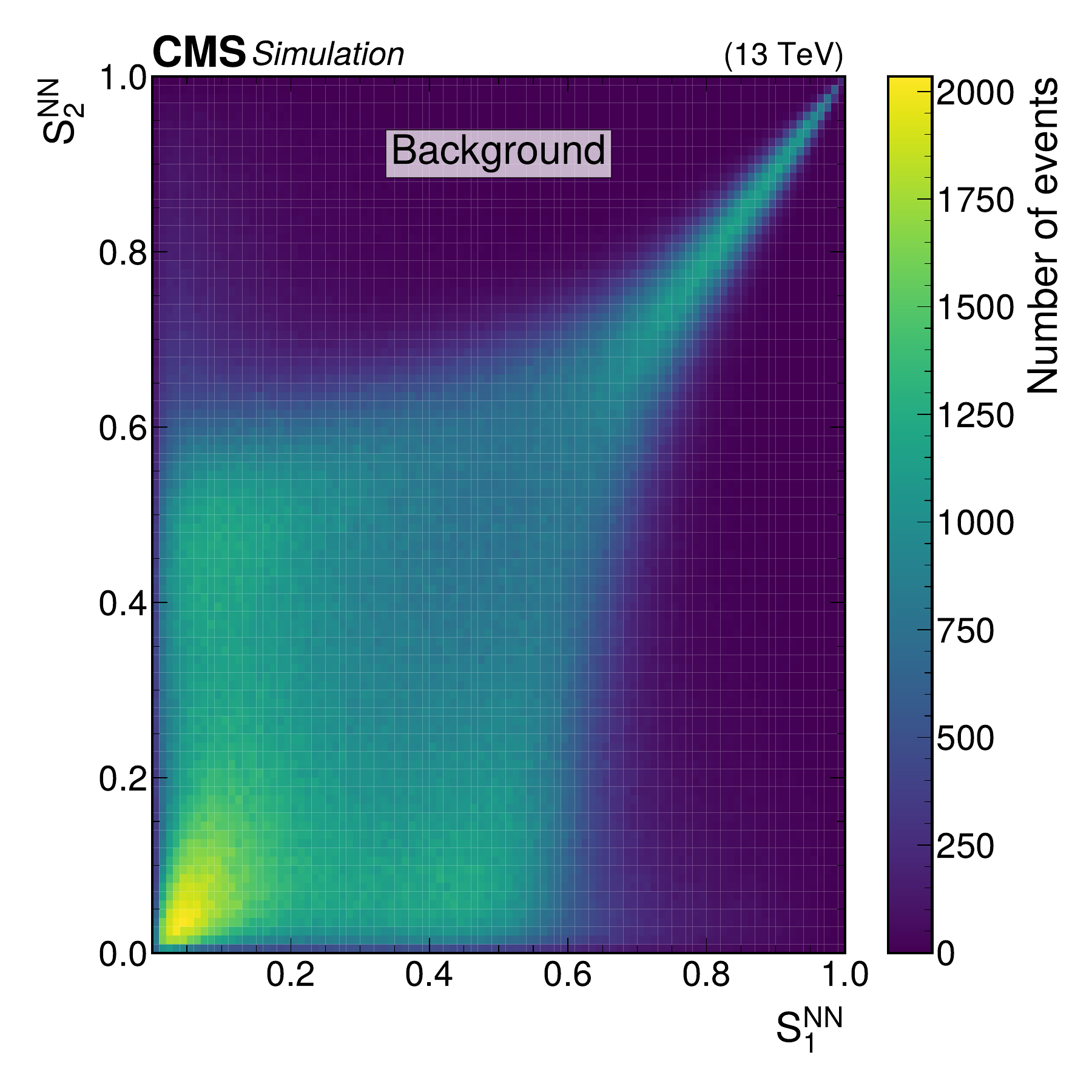}
        \includegraphics[width=0.49\textwidth]{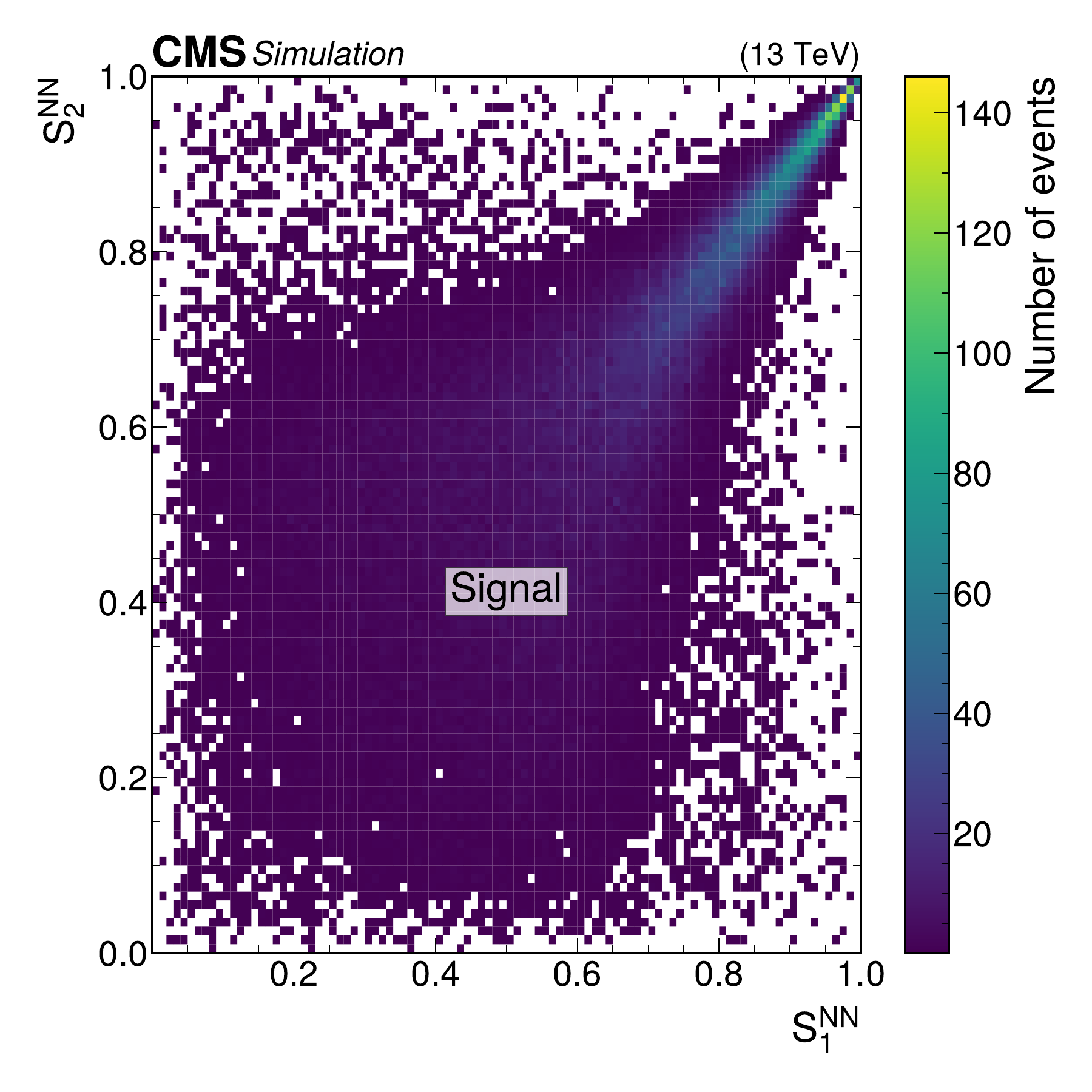}
    \caption{Example of the ``hard edge'' failure mode for background events (\cmsLeft) and signal events (\cmsRight). 
    If the distance correlation loss is evaluated only with background-labeled events, then the interplay between the DisCo and BCE losses for the classifiers can yield a solution where decorrelation is only achieved in a subregion of the two-dimensional plane.
    White bins contain no events.}
    \label{fig:hardEdge}
\end{figure}

Another failure mode arises when considering whether to apply the full loss function to both signal and background. 
One could apply the DisCo loss only to events from the background data sets, which might allow signal to concentrate more strongly in the \Areg region without the DisCo loss forcing it to disperse into surrounding regions. 
However, such a configuration leads to hard edges in the background distribution where very few events populate the regions beyond 0.5 for either discriminant, as shown in \Fig{hardEdge}. 
Beyond these hard edges, the small fraction of background events are clearly no longer decorrelated. 
This behavior indicates that the penalty for misclassifying a signal-like event as background is larger than the penalty for correlation imposed by the DisCo loss. 
Thus, enforcing low distance correlation for both signal and background ensures that solutions do not converge to this failure mode.

While careful monitoring and tuning of hyperparameters can help to avoid these failure modes and otherwise poor performance, such an effort can be exceedingly time consuming. 
To mitigate this burdensome trial-and-error procedure, the MDMM approach introduced in \Sec{mdmm} directly modifies the loss function and training procedure to automate the process of finding the optimal regions of hyperparameter space. 

\section{Results}
\label{sec:results}

Two networks are trained to demonstrate the \discotec method for the physics models described in Section~\ref{sec:susyexample}. 
The first model is trained using the random grid search approach---referred to as the $\lambda$ method---for hyperparameter optimization, with validation results presented in Section~\ref{sec:closure}.
The second model is trained with the same architecture but using MDMM, and the validation results are shown in Section~\ref{sec:mdmm_app_res}.

\subsection{Results from the \texorpdfstring{$\lambda$}{lambda} method}
\label{sec:closure}

The model performance is quantified by assessing how the nonclosure varies across the two-dimensional plane constructed from the two network output discriminants using simulated events.
This is equivalent to checking that the background estimate from the ABCD method is valid and robust.
In the ideal scenario, the success condition $\NApred \approx \NAreg$ would be exactly satisfied for any choice of boundaries defining the ABCD regions.
However, the network must compromise between achieving statistical independence of the two outputs and separating signal from background across the plane;
therefore, the prediction may only match the observation in most but not all of the plane.

\begin{figure}[b]
    \centering
        \includegraphics[width=0.49\textwidth]{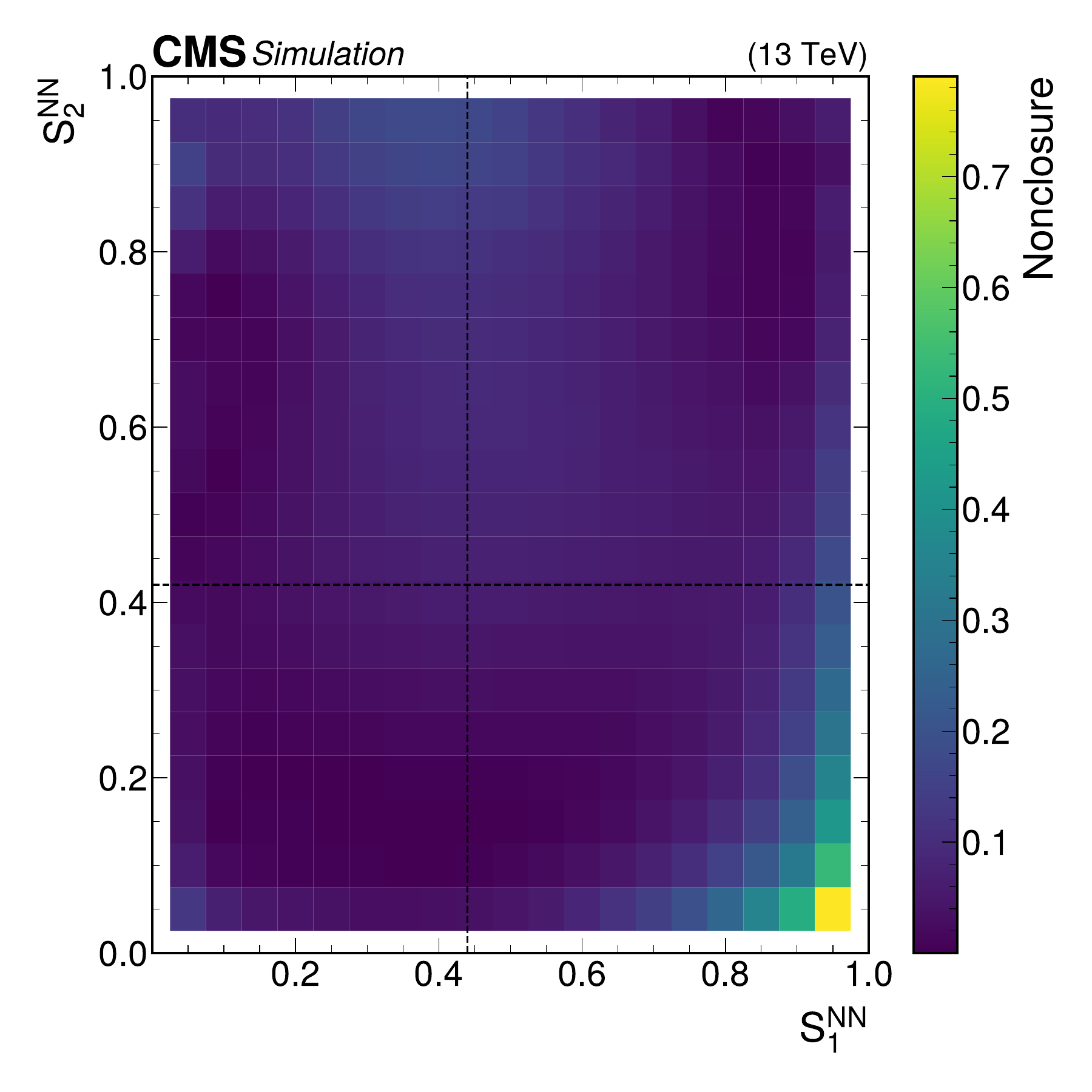}
        \includegraphics[width=0.49\textwidth]{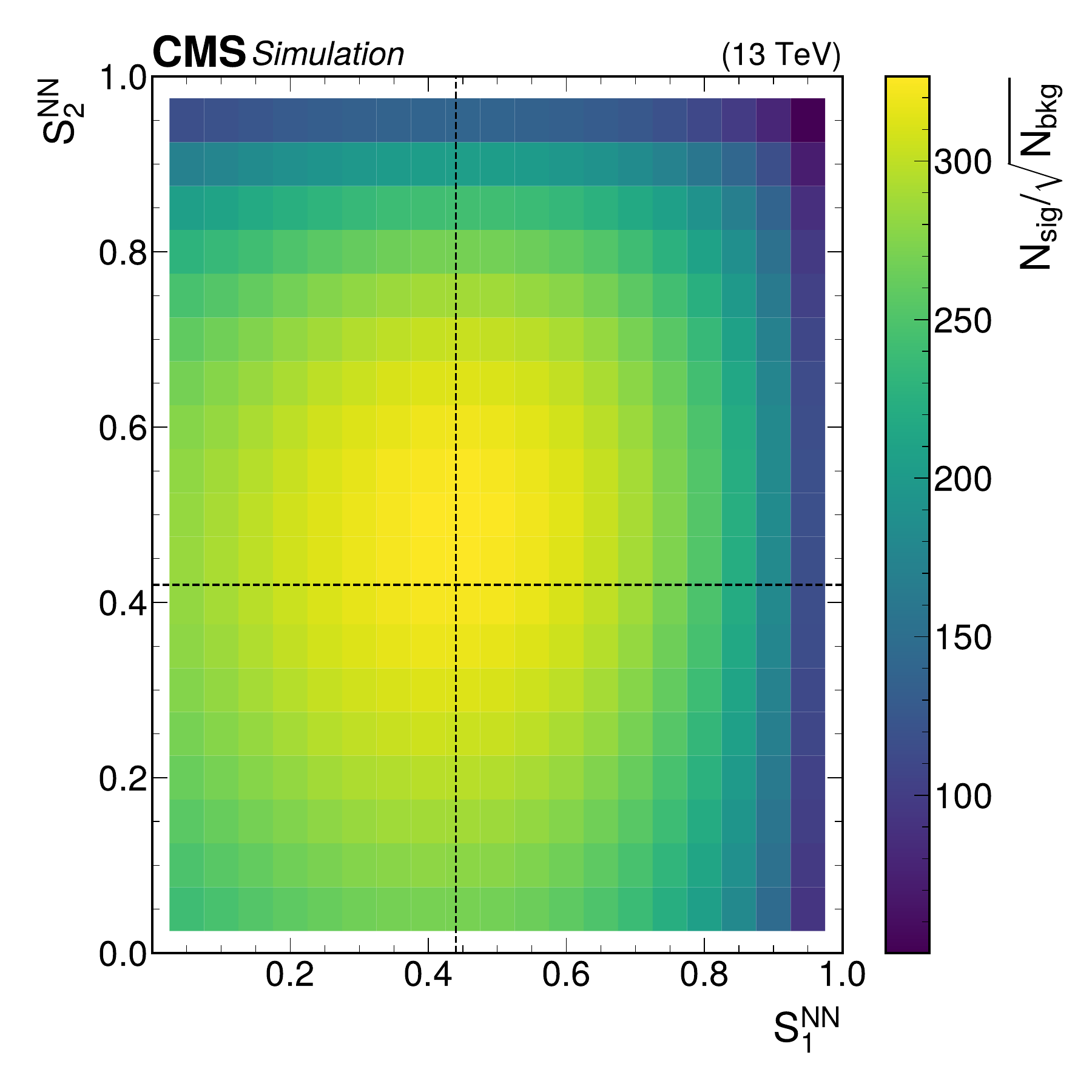}
    \caption{The nonclosure (\cmsLeft) and raw statistical significance ($\Nsig/\sqrt{\smash[b]{\Nbkg}}$) (\cmsRight) as a function of the placement of the ABCD region boundaries for the stealth SUSY training set.
    The dashed black lines represent the optimal boundaries for both nonclosure and statistical significance. Low nonclosure values, below ${\approx}20\%$, are obtained over most of the plane.}
    \label{fig:signNonClosure2D}
\end{figure}

\Figur{signNonClosure2D} shows the nonclosure and raw statistical significance estimates for different boundaries using the simulated test sample, integrated over all \njets categories. 
The robustness of the trained \discotec model can be inferred from the uniform and small nonclosure across the majority of the ABCD plane.
Large nonclosure does occur at the peripheries of the plane where one or more of the ABCD regions become small enough to be dominated by statistical fluctuations from the number of background events approaching zero.
To account for residual nonclosure from the final choice of ABCD boundaries, a correction factor can be applied, or a systematic uncertainty can be assigned. 

The determination of the optimal boundaries is the final step to use the ABCD method.
If the model shows near zero nonclosure for any choice of boundaries, the optimal boundaries can be selected as those producing the largest raw statistical significance.
However, in cases where there is still some nonnegligible nonclosure, it is recommended to include this as a systematic uncertainty during the significance optimization, as shown in \Equ{sigWithNonClosure}. 
Finally, in cases where other systematic uncertainties are expected to have a large effect on the results, pseudo-data studies can be performed for a grid of possible boundary choices. 
This entails running the full statistical interpretation to choose the best boundaries, accounting for all aspects of the analysis.
While this final method is the most robust, it is also the most resource intensive. 
Thus, the costs and benefits of each optimization must be weighed to decide on the appropriate approach.

For the case study shown above, a grid of boundary values are selected between 0.4 and 0.9 in both discriminant dimensions with a spacing of 0.02.
Each bin edge choice is used to derive the background estimate for each analysis category, and a pseudo-data fit is conducted using the full set of systematic uncertainties. 
The statistical significance as calculated by the fit is used as the metric for determining the optimal bin edge choice. 
The optimal bin edges for this particular model are determined to be $(\snni{1}, \snni{2}) = (0.44, 0.42)$.
The slight deviation of these boundaries from the area with the largest raw statistical significance can be attributed to the nonclosure systematic uncertainty in the $\ttjets$ background estimation.

\subsection{Results from MDMM}
\label{sec:mdmm_app_res}

The same network architecture and training procedure for the results in \Sec{closure} are used for the training applying the MDMM approach to the \discotec method.
The only differences are that the loss function is now defined as in \Equ{lagrangian} with the added damping terms from \Equ{quadraticConstraints}, and that the \PYTORCH~\cite{Paszke:2019} package is used to define and train the network.

\begin{figure}
    \centering
        \includegraphics[width=0.49\textwidth]{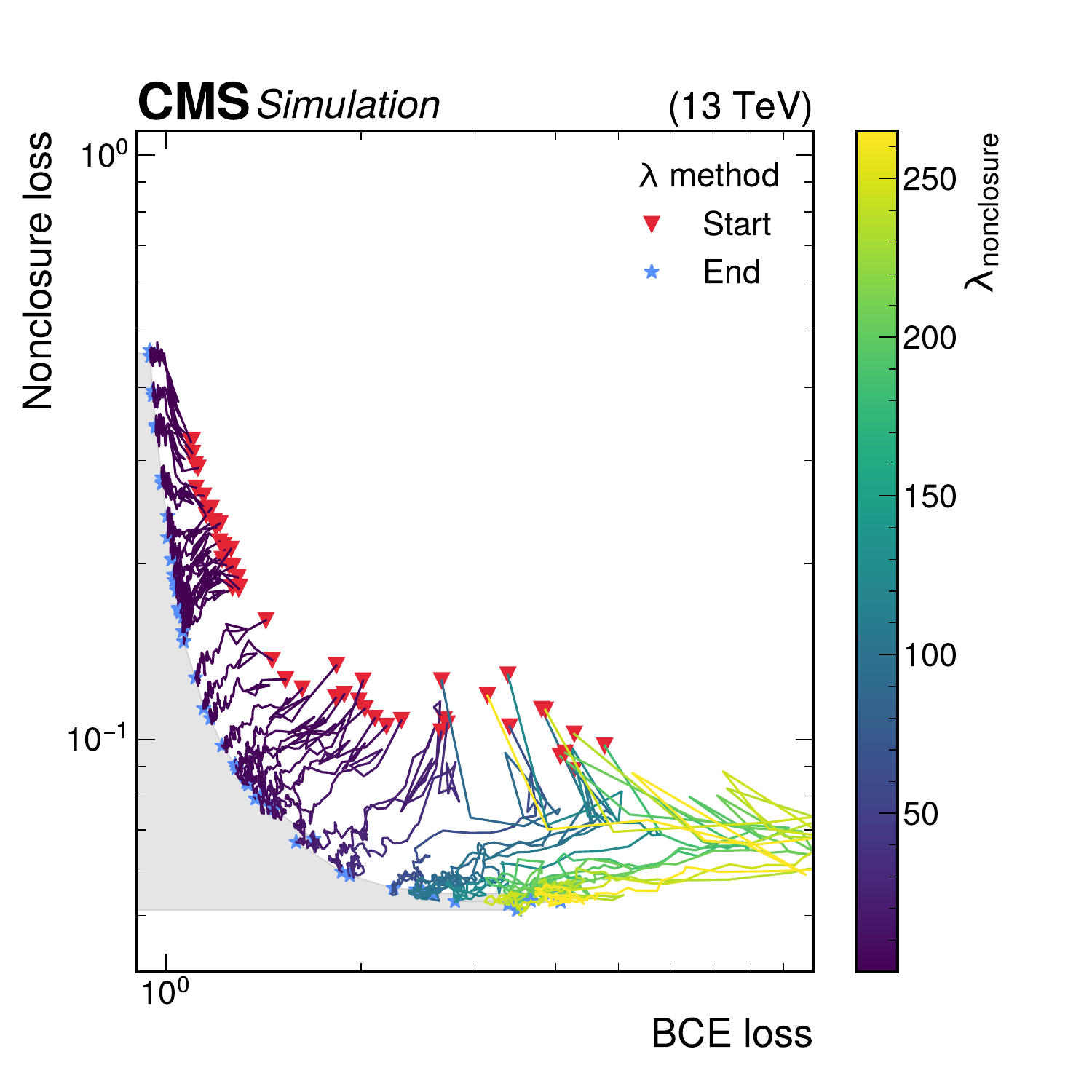}
        \includegraphics[width=0.49\textwidth]{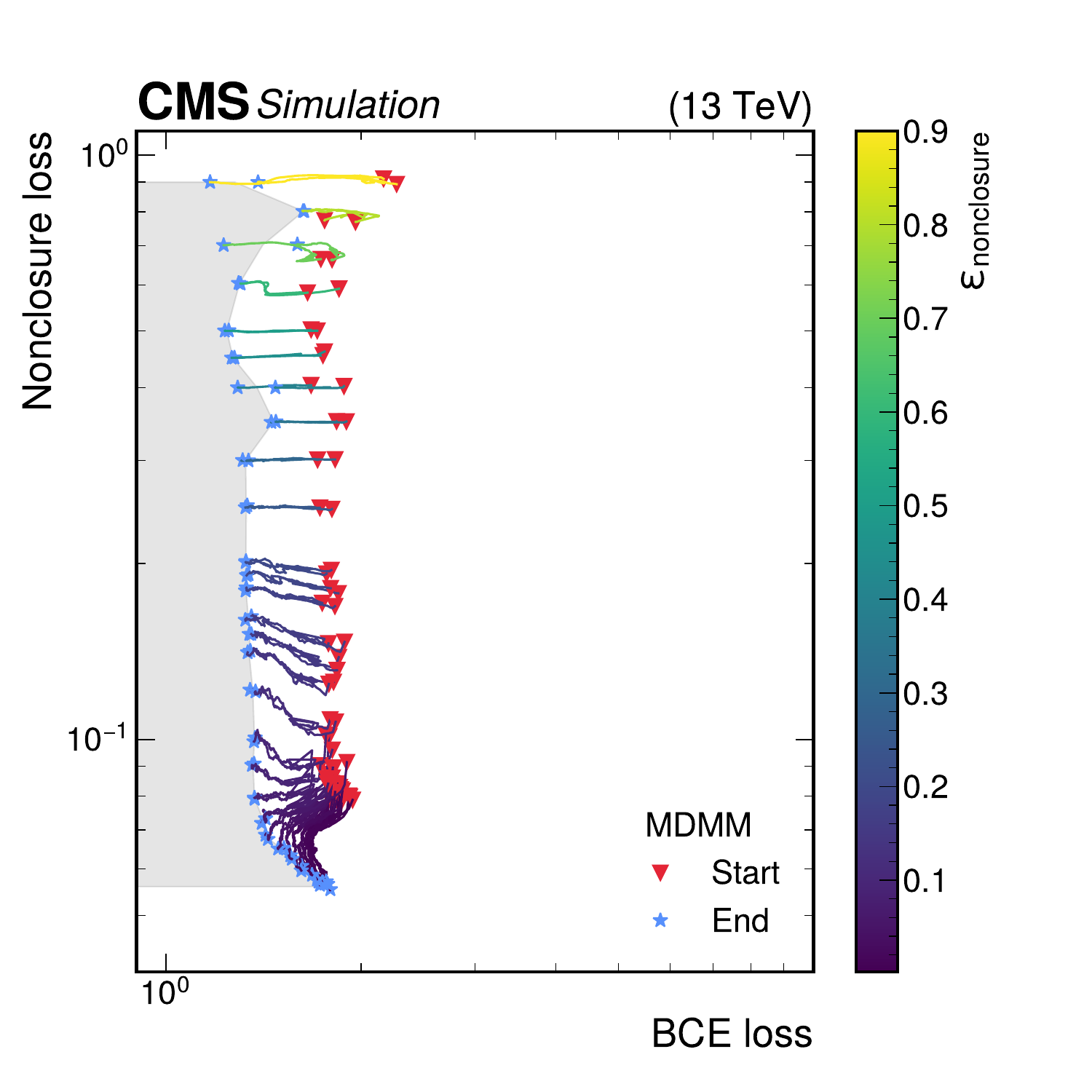}
    \caption{Scans of the Pareto front in the \LClosure vs. \LBCE plane with the $\lambda$ (\cmsLeft) and MDMM (\cmsRight) methods.
    All trainings are conducted with \PYTORCH. 
    Each line shows the training of a model with different values of the $\lambda$ or $\epsilon$ parameters, respectively.}
    \label{fig:pareto_mdmm}
\end{figure}

Figure~\ref{fig:pareto_mdmm} shows scans of the loss values with and without MDMM over the course of multiple trainings, for which \PYTORCH is employed in both cases.
The training with MDMM varies \epsClosure, while the training without MDMM varies \lamClosure.
The MDMM approach is able to explore the entire Pareto front, while the $\lambda$ method is restricted to the region in which it is convex.
Because the shape of the Pareto front is not known a priori, the $\lambda$ method is not guaranteed to probe the whole set of possible solutions; this problem is exacerbated when constraining multiple loss terms. 
In contrast, MDMM allows for a comprehensive hyperparameter scan along the Pareto front, independent of its shape.
In the stealth SUSY application presented in this paper, most of the Pareto front happens to be convex.
Thus, the $\lambda$ method is able to explore the main region of interest. 

However, MDMM is still beneficial to set physically meaningful constraints on all subordinate loss terms.
For example, including a constraint on the nonclosure loss of $\epsClosure = 0.1$ results in a solution that exhibits 10\% nonclosure on average. 
In the $\lambda$ method, a choice of $\lamClosure = 0.1$ has no clear physical meaning, but only determines the relative importance of \LClosure compared to the other loss components. 
By using MDMM to directly set an upper bound on the allowed nonclosure via the loss constraint, fewer trainings are needed to reach the desired optimum.

\begin{figure*}[htp]
    \centering
        \includegraphics[width=0.98\textwidth]{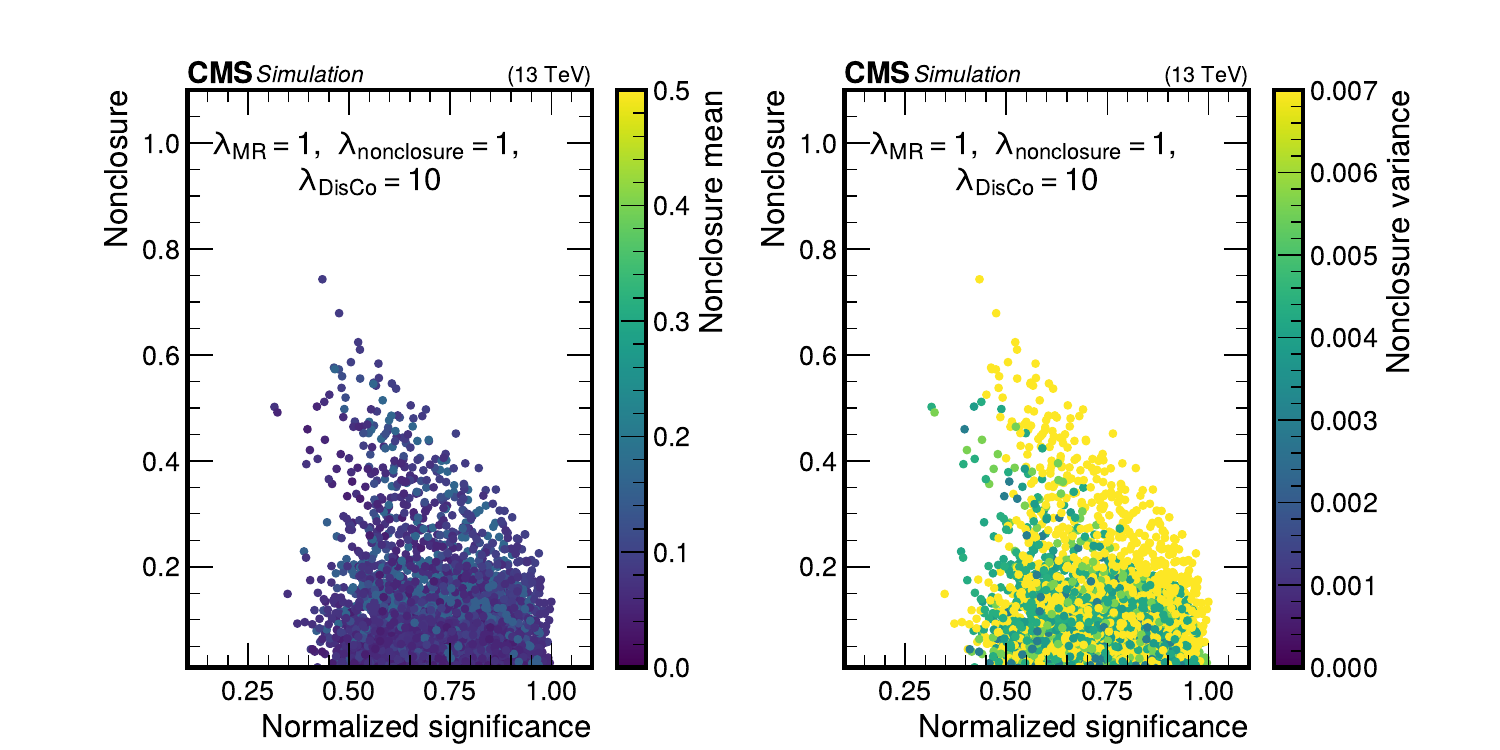}
        \includegraphics[width=0.98\textwidth]{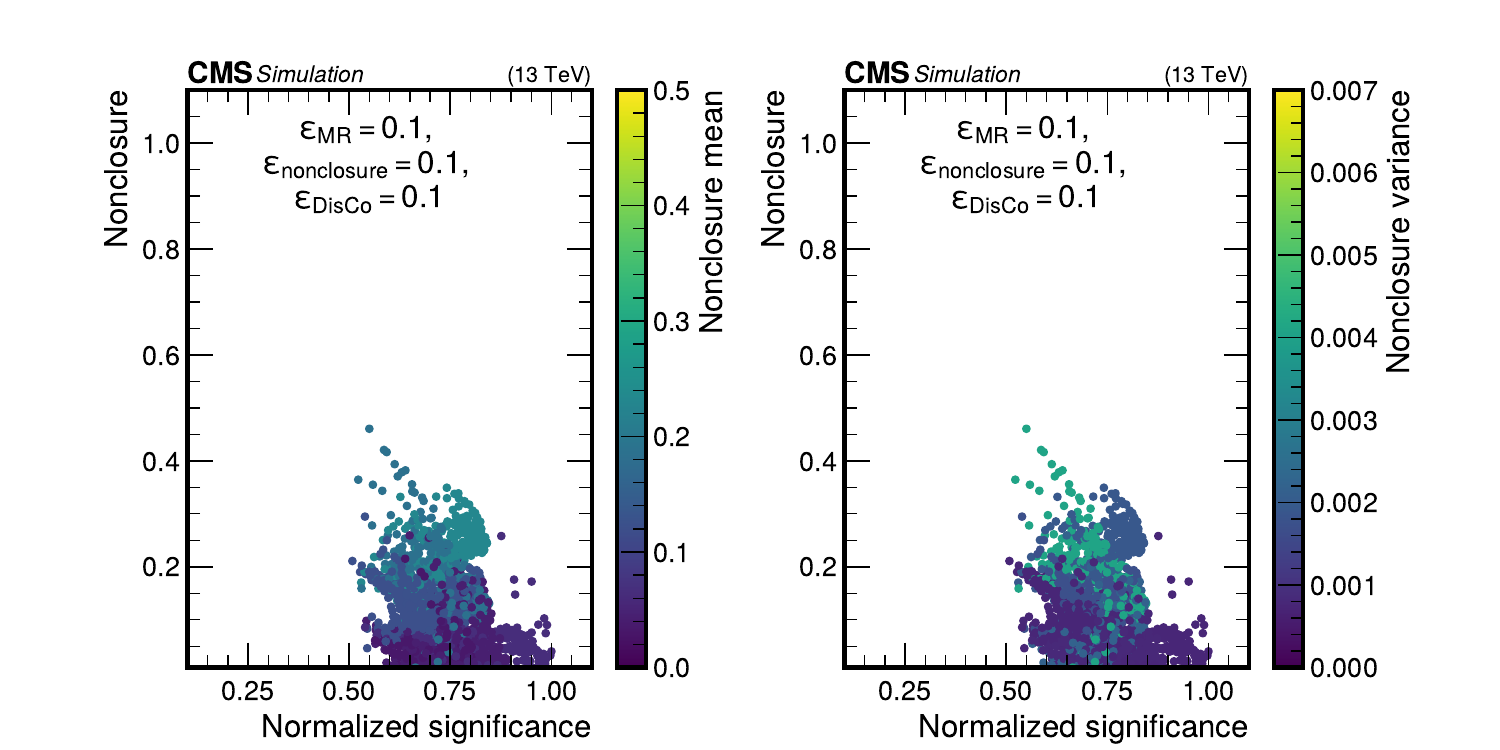}
    \caption{The significance and nonclosure values from scanning the ABCD boundaries, with training using the $\lambda$ method (upper) and MDMM (lower).
    Multiple trainings with different initial weights and biases are shown for each of the two training approaches.
	Each point represents a given training and a given choice of boundaries; the color of the point represents the nonclosure mean (left) and variance (right) from all boundary choices for the training used for that point.
    }
    \label{fig:lambdaVsMDMM}
\end{figure*}

The MDMM approach also results in more robust and stable predictions from the ABCD method.
This is examined in the significance-closure plane for fixed $\epsilon$ or $\lambda$, using MDMM and the $\lambda$ method, respectively. 
In this case, trainings are conducted with constraints on both the DisCo and nonclosure loss terms.
Multiple trainings are conducted for each method with different random sequences used in each training, including the initialization of the network weights.

The results for both methods are shown in \Fig{lambdaVsMDMM} and can be described in terms of two criteria.
First, given a set of constraint values, a network is said to be ``iteratively robust'' if the majority of trainings result in the same nonclosure and significance, as computed in \Equ{sigWithNonClosure}.
This is identified in \Fig{lambdaVsMDMM} by comparing the mean nonclosure from each training.
Second, a network is said to be ``closure robust'' if, in a given training, the nonclosure variance between different boundaries in a given training is low. 
This implies a low nonclosure variance for each of the trainings.
Iteratively robust networks require fewer training iterations to converge, while closure robust networks result in generally smaller systematic uncertainties associated with nonclosure. 

For this example, the trainings using the MDMM approach are more robust in both categories defined above. 
The MDMM trainings yield a tighter clustering of trainings in the nonclosure and significance plane, meaning that these trainings are more robust to initial conditions than trainings using the $\lambda$ method. 
Therefore, it is easy to find a training that provides the desired outcome by setting the proper constraint on the nonclosure with a small number of iterations.
Additionally, the MDMM approach results in more closure robust trainings, as indicated by the low nonclosure variances.
This implies that MDMM is more likely to produce trainings with broadly good closure.

The overall performance of trainings with MDMM and the $\lambda$ method are similar. 
This is in part due to the convex nature of the Pareto front for this specific case study. 
However, since the shape of the Pareto front is not known a priori, it is recommended to use MDMM with constraints tailored to the needs of the analysis. 
This way, the network is more likely to converge to an optimal solution with robust closure, compared to the $\lambda$ method.
Additionally, to converge on the final model in the case presented here, ${\approx}1000$ iterations of the $\lambda$ training procedure were needed, while only ${\approx}10$ with MDMM.
Thus, MDMM enables intuitive constraints on loss components and quicker convergence.

\section{Validation with CMS data}
\label{sec:data_validation}

To apply the \discotec method to proton-proton collision data, the performance of the method should transfer from simulated events used for training to real data collision events, both regarding signal vs. background separation and the independence of the two discriminants.
Any differences in behavior between simulation and observed data must be taken into account to draw correct conclusions about the possible presence of signal. 
It is thus important to evaluate the network's behavior in a VR in observed data to assess appropriate corrections and systematic uncertainties. 

A typical physics analysis using the \discotec method would define a ``target region'' (TR) using event selection requirements based on the signature of interest.
The \discotec method would be applied to provide the background estimation in the subset of the TR defined as the SR.
In some physics analyses, it is possible to define VRs via an orthogonal set of event selection requirements that retains the same kinematic behavior of the key observables used in the NN training.
When such a region can be found, the validation of the \discotec model straightforwardly proceeds by directly assessing the closure within that region.

When it is not feasible to define such a region with orthogonal selection requirements, it is possible to use the nature of the ABCD method itself to define VRs within the TR. This is the approach used in the case study. 
Because of the possibility of signal events being present in the collision data, the full two-dimensional ABCD plane in the TR cannot be safely used to check the level of nonclosure.
However, the \Breg, \Creg, and \Dreg CRs have a smaller signal fraction than the SR \Areg, by construction in the \discotec method.
Those regions can therefore be used to evaluate the nonclosure behavior.
The combination of these three regions spans the full range of values for each NN discriminant, allowing for a near-full validation of the method directly in data.
The robustness of the method is determined by two observations: similar levels of nonclosure in simulation and observed data, and slowly varying nonclosure across the full plane (including the \Areg region) in simulation.

\begin{figure*}[htp]
\centering
\includegraphics[width=0.75\textwidth]{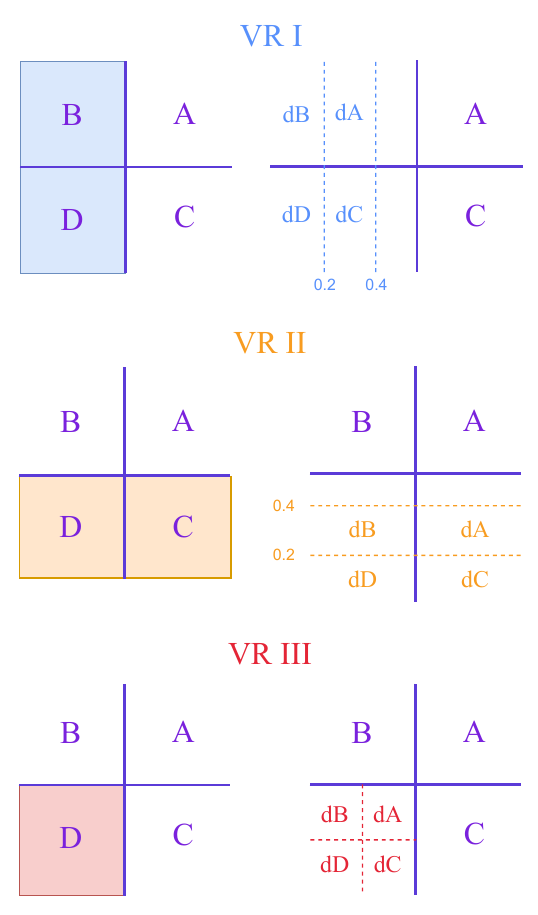}
    \caption{Visual representations of the three VRs (VR I, VR II, and VR III).
    The solid blue lines indicate the bin boundaries from the full ABCD plane.
    The dashed lines represent the starting points of the ABCD subregion edges that are moved during the validation procedure, as described in the text. 
    These lines divide the VRs into regions dA, dB, dC, and dD; the latter three are used to predict the number of events in dA, following the ABCD prescription.
\label{fig:val_cartoon}}
\end{figure*}

This test is performed in the stealth SUSY case study, using the model trained in \Sec{closure}.
Full details on the data sets used are provided in Ref.~\cite{SUS-23-001}. 
The first validation region (VR I) is created using the left side of the ABCD plane, roughly corresponding to the combination of the {\Breg} and \Dreg regions.
The right-most (outer) boundary of VR I is placed at $\snni{1}=0.4$, and a sub-ABCD plane is defined by placing another boundary at $\snni{1}=0.2$. The inner boundary in \snni{2} is placed at the same value as in the full ABCD plane.
Then, the boundaries in \snni{1} are iteratively increased toward the boundaries for the full ABCD plane.
The outermost boundary of the VR is increased in steps of 0.05 for each iteration.
At the same time, the inner boundary in \snni{1} is changed in linear steps such that it arrives at the optimal bin edge for the full ABCD plane when the outer boundary reaches $\snni{1}=1.0$.  
In each iteration, the nonclosure is measured in both simulation and observed data.
The stepping procedure is stopped when the expected signal contribution becomes too large, defined here as 5\% signal fraction in the analogue of the \Areg region in the sub-ABCD plane.
A similar procedure is followed for the second validation region (VR II) using the lower side of the ABCD plane, roughly corresponding to the combination of the \Creg and \Dreg regions.
The third validation region (VR III) uses the lower left corner of the plane, with the starting point of the procedure being the full \Dreg region.
The sub-ABCD plane is created by bisecting the \Dreg region on both axes.
Then, all boundaries are increased towards those defining the full ABCD plane simultaneously, as described for the stepping procedure for \snni{1} for VR I.  
A graphical representation of these validation region definitions is shown in \Fig{val_cartoon}. 

\begin{figure*}[tp]
\centering
\includegraphics[width=0.90\textwidth]{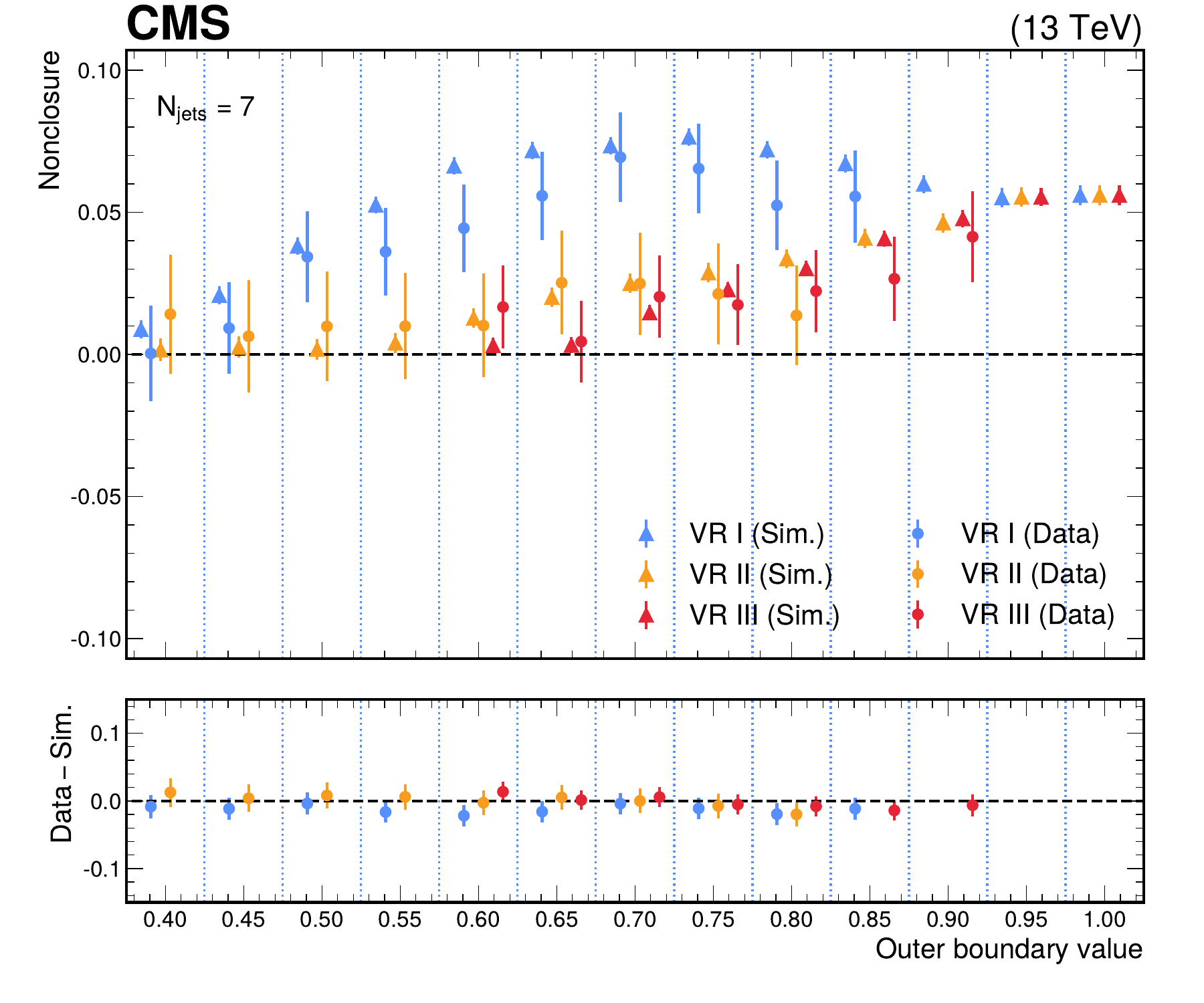}
\caption{The nonclosure in simulation and observed data when iteratively increasing the sub-ABCD region boundaries as described in the text, for VR I, VR II, and VR III.
    The scans of VR I and VR II start at the outer boundary value of 0.4, while the scan for VR III starts at the boundary value of 0.6.
    Observed data are shown only for boundary values where the signal contamination is below 5\% in the analogue of the A region.
\label{fig:data_validation}}
\end{figure*}

In all cases, good agreement is achieved between data and simulation as observed in \Fig{data_validation}. Any small disagreements can be taken into account as a systematic uncertainty in the analysis.
As reported in Ref.~\cite{SUS-23-001}, the systematic uncertainty determined via this VR approach for the stealth SUSY search is 3--15\%, depending on the analysis category. 
The \discotec method improves the signal sensitivity compared to the previous search in Ref.~\cite{CMS:2021knz}: for example, the expected upper limit on the signal cross section is improved by a factor of ${\approx}2$ for top squark masses between 300 and 450\GeV. 

\section{Summary and outlook}
\label{sec:discussion}

The ABCD background estimation technique uses three control regions, defined by boundaries in the space of two statistically independent variables, to predict the background in the signal region.
There has been ongoing interest in automating this technique with machine learning, such as the \disco method~\cite{Kasieczka:2020pil}, which uses the distance correlation (DisCo) metric to train a neural network to produce two independent discriminants.
In this paper, we introduce the \disco training enhanced with closure (\discotec) method, which improves this automation by adding a novel differentiable loss term. This term quantifies the deviation in the background prediction (``nonclosure''), more directly addressing the primary objective of the ABCD technique.

The result, from a case study using a high-energy physics data set with a simulated stealth supersymmetry signal, is a pair of decorrelated discriminants with strong signal-background separation, reducing the systematic bias in the background estimation and improving the significance for a potential observation of physics beyond the standard model.
Strategies for validating the method in proton-proton collision data and deriving applicable systematic uncertainties were presented. 
These achievements represent the first application of such methods to observed collider data and promising steps towards a generalized strategy to automate background estimation in high-energy physics.

Additionally, an alternative approach to train the \discotec neural network was evaluated, using the modified differential method of multipliers (MDMM)~\cite{Platt:1987}.
The MDMM approach promotes the relative weights of the additional loss terms to learnable parameters, specifies constraints on each additional loss term directly, and guarantees the convergence of the training.
This method provides further benefits when applied to \discotec by more quickly and directly determining optimal regions in hyperparameter space, without requiring exhaustive manual searches.
This also constitutes the first use of MDMM with collider data.

Several potential modifications or extensions to the loss function could further improve the performance of the \discotec method.
For example, incorporating a normalized measure of the signal yield in the control regions as an additional loss term could directly constrain the allowed contamination~\cite{Kasieczka:2020pil}.
This modification could improve the network's sensitivity to the signal while simplifying validation by reducing high signal contamination in certain regions.
Another promising extension involves augmenting the nonclosure loss term based on the extended ABCD method~\cite{Choi:2019mip}, which uses additional control regions to mitigate nonclosure effects from minor correlations between the discriminants.
By following the approach employed here, the extended nonclosure loss can be made differentiable by relaxing hard boundaries using sigmoid functions.

In conclusion, the combination of the \discotec method and MDMM represents another step forward in the optimization and automation of the traditional ABCD background estimation method.
This technique has promising applications to a wide variety of high-energy physics analyses and the potential for further refinement in the future.

\begin{acknowledgments}
We congratulate our colleagues in the CERN accelerator departments for the excellent performance of the LHC and thank the technical and administrative staffs at CERN and at other CMS institutes for their contributions to the success of the CMS effort. In addition, we gratefully acknowledge the computing centers and personnel of the Worldwide LHC Computing Grid and other centers for delivering so effectively the computing infrastructure essential to our analyses. Finally, we acknowledge the enduring support for the construction and operation of the LHC, the CMS detector, and the supporting computing infrastructure provided by the following funding agencies: SC (Armenia), BMBWF and FWF (Austria); FNRS and FWO (Belgium); CNPq, CAPES, FAPERJ, FAPERGS, and FAPESP (Brazil); MES and BNSF (Bulgaria); CERN; CAS, MoST, and NSFC (China); MINCIENCIAS (Colombia); MSES and CSF (Croatia); RIF (Cyprus); SENESCYT (Ecuador); ERC PRG, RVTT3 and MoER TK202 (Estonia); Academy of Finland, MEC, and HIP (Finland); CEA and CNRS/IN2P3 (France); SRNSF (Georgia); BMBF, DFG, and HGF (Germany); GSRI (Greece); NKFIH (Hungary); DAE and DST (India); IPM (Iran); SFI (Ireland); INFN (Italy); MSIP and NRF (Republic of Korea); MES (Latvia); LMTLT (Lithuania); MOE and UM (Malaysia); BUAP, CINVESTAV, CONACYT, LNS, SEP, and UASLP-FAI (Mexico); MOS (Montenegro); MBIE (New Zealand); PAEC (Pakistan); MES and NSC (Poland); FCT (Portugal); MESTD (Serbia); MICIU/AEI and PCTI (Spain); MOSTR (Sri Lanka); Swiss Funding Agencies (Switzerland); MST (Taipei); MHESI and NSTDA (Thailand); TUBITAK and TENMAK (Turkey); NASU (Ukraine); STFC (United Kingdom); DOE and NSF (USA).

\hyphenation{Rachada-pisek} Individuals have received support from the Marie-Curie program and the European Research Council and Horizon 2020 Grant, contract Nos.\ 675440, 724704, 752730, 758316, 765710, 824093, 101115353, 101002207, 101001205, and COST Action CA16108 (European Union); the Leventis Foundation; the Alfred P.\ Sloan Foundation; the Alexander von Humboldt Foundation; the Science Committee, project no. 22rl-037 (Armenia); the Fonds pour la Formation \`a la Recherche dans l'Industrie et dans l'Agriculture (FRIA-Belgium); the Beijing Municipal Science \& Technology Commission, No. Z191100007219010, the Fundamental Research Funds for the Central Universities, the Ministry of Science and Technology of China under Grant No. 2023YFA1605804, and the Natural Science Foundation of China under Grant No. 12061141002 (China); the Ministry of Education, Youth and Sports (MEYS) of the Czech Republic; the Shota Rustaveli National Science Foundation, grant FR-22-985 (Georgia); the Deutsche Forschungsgemeinschaft (DFG), among others, under Germany's Excellence Strategy -- EXC 2121 ``Quantum Universe" -- 390833306, and under project number 400140256 - GRK2497; the Hellenic Foundation for Research and Innovation (HFRI), Project Number 2288 (Greece); the Hungarian Academy of Sciences, the New National Excellence Program - \'UNKP, the NKFIH research grants K 131991, K 133046, K 138136, K 143460, K 143477, K 146913, K 146914, K 147048, 2020-2.2.1-ED-2021-00181, TKP2021-NKTA-64, and 2021-4.1.2-NEMZ\_KI-2024-00036 (Hungary); the Council of Science and Industrial Research, India; ICSC -- National Research Center for High Performance Computing, Big Data and Quantum Computing and FAIR -- Future Artificial Intelligence Research, funded by the NextGenerationEU program (Italy); the Latvian Council of Science; the Ministry of Education and Science, project no. 2022/WK/14, and the National Science Center, contracts Opus 2021/41/B/ST2/01369 and 2021/43/B/ST2/01552 (Poland); the Funda\c{c}\~ao para a Ci\^encia e a Tecnologia, grant CEECIND/01334/2018 (Portugal); the National Priorities Research Program by Qatar National Research Fund;  MICIU/AEI/10.13039/501100011033, ERDF/EU, "European Union NextGenerationEU/PRTR", and Programa Severo Ochoa del Principado de Asturias (Spain); the Chulalongkorn Academic into Its 2nd Century Project Advancement Project, and the National Science, Research and Innovation Fund via the Program Management Unit for Human Resources \& Institutional Development, Research and Innovation, grant B39G670016 (Thailand); the Kavli Foundation; the Nvidia Corporation; the SuperMicro Corporation; the Welch Foundation, contract C-1845; and the Weston Havens Foundation (USA).    
\end{acknowledgments}

\bibliography{auto_generated}
\cleardoublepage \appendix\section{The CMS Collaboration \label{app:collab}}\begin{sloppypar}\hyphenpenalty=5000\widowpenalty=500\clubpenalty=5000\input{MLG-23-003-public-authorlist.tex}\end{sloppypar}
\end{document}

%% file: MLG-23-003-public-authorlist.tex
\cmsinstitute{Yerevan Physics Institute, Yerevan, Armenia}
{\tolerance=6000
A.~Hayrapetyan, V.~Makarenko\cmsorcid{0000-0002-8406-8605}, A.~Tumasyan\cmsAuthorMark{1}\cmsorcid{0009-0000-0684-6742}
\par}
\cmsinstitute{Institut f\"{u}r Hochenergiephysik, Vienna, Austria}
{\tolerance=6000
W.~Adam\cmsorcid{0000-0001-9099-4341}, J.W.~Andrejkovic, L.~Benato\cmsorcid{0000-0001-5135-7489}, T.~Bergauer\cmsorcid{0000-0002-5786-0293}, K.~Damanakis\cmsorcid{0000-0001-5389-2872}, M.~Dragicevic\cmsorcid{0000-0003-1967-6783}, C.~Giordano, P.S.~Hussain\cmsorcid{0000-0002-4825-5278}, M.~Jeitler\cmsAuthorMark{2}\cmsorcid{0000-0002-5141-9560}, N.~Krammer\cmsorcid{0000-0002-0548-0985}, A.~Li\cmsorcid{0000-0002-4547-116X}, D.~Liko\cmsorcid{0000-0002-3380-473X}, I.~Mikulec\cmsorcid{0000-0003-0385-2746}, J.~Schieck\cmsAuthorMark{2}\cmsorcid{0000-0002-1058-8093}, R.~Sch\"{o}fbeck\cmsAuthorMark{2}\cmsorcid{0000-0002-2332-8784}, D.~Schwarz\cmsorcid{0000-0002-3821-7331}, M.~Shooshtari, M.~Sonawane\cmsorcid{0000-0003-0510-7010}, W.~Waltenberger\cmsorcid{0000-0002-6215-7228}, C.-E.~Wulz\cmsAuthorMark{2}\cmsorcid{0000-0001-9226-5812}
% ...
\par}
\vskip\cmsinstskip
\dag:~Deceased\\
$^{1}$Also at Yerevan State University, Yerevan, Armenia\\
$^{2}$Also at TU Wien, Vienna, Austria\\
% ...
